\documentclass[11pt,letterpaper]{article}
\usepackage{jheppub}
\usepackage{subcaption,multirow}
\usepackage{verbatim}
\pdfoutput=1 
\usepackage[T1]{fontenc} 
\usepackage[normalem]{ulem}
\usepackage{color} 
\usepackage{amsfonts,amssymb,amsmath} 
\usepackage{mathtools} 
\usepackage{slashed,braket}       
\usepackage{bm}  
\usepackage{hyperref} 
\usepackage{parskip} 
\usepackage{empheq}
\usepackage{tensor} 
\usepackage{mathrsfs}
\usepackage[utf8]{inputenc}
\usepackage{bbold}
\usepackage{booktabs}
\usepackage{perpage} 
\usepackage{slashed}
\usepackage{float}
\usepackage{color}
\usepackage{braket}
\usepackage[table,xcdraw]{xcolor}
\usepackage[makeroom]{cancel}
\usepackage{empheq}
\usepackage{gensymb}
\usepackage{overpic} 
\usepackage{graphicx}
\usepackage{feynmp-auto,tikz}
\usepackage{tikz-feynman} 
\usepackage{slashed} 
 \definecolor{myred}{rgb}{0.804688, 0.09375, 0.117188}

\DeclareMathAlphabet\mathbfcal{OMS}{cmsy}{b}{n}
\newcommand{\eql}[2]{\begin{equation} \label{eqn:#1} #2 \end{equation}}

\newcommand{\eqnref}[1]{Eq.~(\ref{eqn:#1})}

\newcommand{\secref}[1]{Sec.~\ref{sec:#1}}

\newcommand{\ld}[1]{_{\mathrm{#1}}}

\newcommand{\plz}{\partial_z}
\newcommand{\ora}[1]{\overrightarrow{#1}}
\newcommand{\ola}[1]{\overleftarrow{#1}}

\newcommand{\colmatfi}[1]{\left(\begin{array}{ccccc}#1\end{array}\right)}

\def\c{\chi}
\def\d{\delta}
\def\e{\epsilon}

\def\lpar#1#2#3#4{\rlap{\raise#3\hbox{$\hskip#4#1\left\{\mbox{\phantom{\rule[0mm]{0mm}{#2}}}\right.$}}}
\def\rpar#1#2#3#4{\rlap{\raise#3\hbox{$\hskip#4\left\}#1\mbox{\phantom{\rule[0mm]{0mm}{#2}}}\right.$}}}
\addtocounter{subsection}{1}
\renewcommand{\subsubsection}[1]{
	\addtocounter{subsubsection}{1}
	\par\nobreak
	\medskip
	\nobreak
	\noindent{\it \thesubsubsection.  #1 }
	\par\nobreak\medskip\nobreak
}

\title{Continuum Naturalness}

\author[a]{Csaba Cs\'aki,}
\author[a,b]{Gabriel Lee,}
\author[b,c]{Seung J. Lee,}
\author[a]{Salvator Lombardo,}
\author[a]{and Ofri Telem}

\emailAdd{csaki@cornell.edu}
\emailAdd{gabr.lee@cornell.edu}
\emailAdd{sjjlee@korea.edu}
\emailAdd{sdl88@cornell.edu}
\emailAdd{t10ofrit@gmail.com}

\affiliation[a]{Laboratory for Elementary Particle Physics, Cornell University, Ithaca, NY 14853, USA}
\affiliation[b]{Department of Physics, Korea University, Seoul 02841, Republic of Korea}
\affiliation[c]{School of Physics, Korea Institute for Advanced Study, Seoul 130-722, Korea}

\abstract{
We present a novel class of composite Higgs models in which the top and gauge partners responsible for cutting off the Higgs quadratic divergences form a continuum. The continuum states are characterized by their spectral densities, which should have a finite gap for realistic models. We present a concrete example based on a warped extra dimension with a linear dilaton, where this finite gap appears naturally. We derive the spectral densities in this model and calculate the full Higgs potential for a phenomenologically viable benchmark point, with percent level tuning. The continuum top and gauge partners in this model evade all resonance searches at the LHC and yield qualitatively different collider signals.}

\begin{document}

\maketitle	

\section{Introduction}

The cornerstone of conventional solutions to the hierarchy problem, e.g. supersymmetry and composite Higgs (CH) \cite{Georgi:1984af,ArkaniHamed:2001nc,Agashe:2004rs,Contino:2010rs,Bellazzini:2014yua,Panico:2015jxa,Csaki:2015hcd,Csaki:2016kln,Csaki:2018muy}, is the existence of new states around the TeV scale. 
The role of these top and gauge partners is to cut off the quadratically divergent radiative corrections to the Higgs potential from the top quark and gauge bosons. 
In recent years, searches at the LHC have placed the naturalness paradigm under pressure by setting significant lower bounds on the masses of top and gauge partners of about 1.2--1.4 TeV \cite{Sirunyan:2018yun,Sirunyan:2018ncp,Sirunyan:2018omb,Aaboud:2018wxv,Aaboud:2018xpj,Aaboud:2018pii} and 2.2--2.5 TeV \cite{Sirunyan:2018hsl,Sirunyan:2018iff,Aaboud:2018ohp}, respectively.

However, many of these searches assume that the top and gauge partners are particles that can be produced on-shell. 
In this paper we introduce a new class of models in which the top and gauge partners are gapped \textit{continuum} states \cite{Georgi:2007ek,Georgi:2007si,Cacciapaglia:2007jq,Cacciapaglia:2008ns}, rather than ordinary \textit{particles}.

The simplest example of a spectrum with gapped continuum modes is that obtained from the finite potential well in standard quantum mechanics (QM). 
While the bound states inside the well form a discrete set, the scattering states form a continuum with energies above the well. 
Another example is a strongly-interacting theory with a critical behavior in the IR that gives rise to a gapped continuum\footnote{For example, see~\cite{Stancato:2008mp,Falkowski:2008fz,Falkowski:2008yr,Bellazzini:2015cgj}.}. 
For example, it is believed that at the bottom of the conformal window of supersymmetric QCD, a gapped continuum can be generated by turning on a squark mass \cite{Cacciapaglia:2007jq}.

In this work we present a CH model in which continuum top and gauge partners arise as the composites of a strong sector with critical behavior in the IR. 
Inspired by the AdS/CFT correspondence~\cite{Maldacena:1997re} and holographic realizations of CH models~\cite{Contino:2003ve,Agashe:2004rs}, our gapped continuum arises from a warped 5D geometry \cite{Randall:1999ee} with a linear dilaton \cite{Aharony:1998ub,Giveon:1999px,Antoniadis:2001sw}. 
The resulting Green's functions have a branch cut starting at a finite gap $\mu$ corresponding to the slope of the linear dilaton, indicating the emergence of a continuum.

Based on this linear dilaton geometry, we construct a fully realistic CH model with partial compositeness \cite{Kaplan:1991dc,Grossman:1999ra,Gherghetta:2000qt,ArkaniHamed:1999dc,Agashe:2004cp} and  gauge-Higgs unification~\cite{Manton:1979kb,Hosotani:1983xw,Antoniadis:2001cv,Kubo:2001zc,vonGersdorff:2002rg,Cacciapaglia:2005da}. 
Similar to the standard warped 5D realizations of CH, our setting involves AdS$_5$ with a UV brane and an IR brane. 
However, in our case, the fifth dimension continues beyond the IR brane to infinity, with a dilaton rising in the deep IR \cite{Batell:2008me,Batell:2008zm,Cabrer:2009we}. 
The other ingredients of the model are identical to standard CH models: a bulk gauge symmetry $SO(5)\times U(1)$ broken to $SO(4)\times U(1)$ on the IR brane and to $SU(2)_L\times U(1)_Y$ on the UV brane; the $A_5$ of $SO(5)/SO(4)$ playing the role of the pseudo-Nambu-Goldstone (pNGB) Higgs; and the Standard Model (SM) fermions and their partners embedded in bulk $SO(5)$ representations. 

The result is a realistic CH model in which the top and gauge partners are all continua, with no BSM resonances within the reach of the LHC. We demonstrate this by focusing on one point in our parameter space, for which we get a realistic Higgs potential (with 1\% tuning) with gaps of about 1--2 TeV.

The paper is structured as follows. 
In Section~\ref{sec:EffAc} we present the effective action for continuum states, and how the properties of the continua are encoded in their spectral densities. 
In Section~\ref{sec:LD} we show how to model gauge and fermion continua in a warped 5D geometry with a linear dilaton. 
We give an intuitive argument for the emergence of a gapped continuum from this geometry, based on an effective Schr\"odinger equation, and then calculate the continuum spectral densities in a procedure inspired by AdS/CFT. 
Using linear dilaton geometry, we construct a realistic CH model with gauge-Higgs unification and continuum top and gauge partners in Section~\ref{sec:fullmod}. 
For the purpose of breaking the bulk $SO(5)$ symmetry, we introduce an IR brane. 
However, the fifth dimension continues beyond the IR brane to a region where the linear dilaton dominates and leads to the gapped continuum. The role of the Higgs is played as usual by the $A_5$ of the $SO(5)/SO(4)$ generators, while the SM Yukawa couplings originate from the jump conditions for the bulk fields on the IR brane.

Our results are summarized in Section~\ref{sec:summary}, while the detailed calculation of the gauge and fermion spectral densities is given in Section~\ref{sec:specden} (and in the appendices). 
To extract the fermion spectral densities, we solve the 5D inhomogeneous equations of motion (EOM) subject to the UV boundary conditions and the IR jump conditions. 
We account for the bulk VEV of the Higgs--$A_5$ by rotating it into the IR jump conditions as usual \cite{Falkowski:2006vi}. 
We diagonalize the resulting $20\times 20$ fermionic Green's function matrix to obtain all of the fermionic spectral densities in our model. The gauge spectral densities are calculated in a similar manner.

In Section~\ref{sec:Higgspot} we calculate the Coleman-Weinberg potential for the Higgs from the spectral densities of our benchmark point in parameter space. 
We obtain a fully realistic Higgs potential, consistent with electroweak precision bounds on $v/f$ and with a tuning of $1\%$, compared to per mille level tuning in a corresponding composite Higgs model with the same IR scale $R'$ and the same choice of bulk representations \cite{Panico:2012uw}. 
Finally, we comment on the phenomenology of continuum partners: the lack of resonances within the reach of the LHC, bounds from the running of $\alpha_s$, and the way to calculate the pair-production cross section for continuum fermions. 
The phenomenology of continuum partners will be explored further in an upcoming work \cite{CLLT}.

\section{Effective Action for Continuum States} \label{sec:EffAc}

The essential ingredients of CH models are the tower of composite top and gauge partners. 
These states cancel the one-loop SM top and gauge contributions to the Higgs potential below the confinement scale of $\Lambda$, which we take to be about 2--3~TeV. 
The main new aspect of the model we present in this paper is that the critical IR dynamics give rise to a \textit{continuum} of top and gauge partners rather than a tower of ordinary particles. 
To study the phenomenology of such continuum top and gauge partners we need to first explain how to write an effective action for these states. 

We will illustrate this by presenting the general effective action for a continuum Weyl fermion. 
We start with the Lagrangian for an ordinary massless left-handed (LH) Weyl fermion $\chi$:
\begin{eqnarray}\label{eq:Weyl}
\mathcal{L}_{\chi}~=~-i\bar{\chi}\bar{\sigma}^\mu p_\mu\chi\, .
\end{eqnarray}
The two-point function in momentum space is simply the inverse of the bilinear term in Eq.~\ref{eq:Weyl},
\begin{eqnarray}
\left<\bar{\chi}\chi\right>~=~\frac{i}{\bar{\sigma}^\mu p_\mu}=\frac{i\sigma^\mu p_\mu}{p^2}\, .
\end{eqnarray}
The Lagrangian for a continuum Weyl fermion generalizes Eq.~\ref{eq:Weyl} by including a momentum-dependent form factor $G(p^2)$:
\begin{eqnarray}\label{eq:WeylCont}
\mathcal{L}^{\text{cont.}}_{\chi}~=~-i\bar{\chi}\frac{\bar{\sigma}^\mu p_\mu}{p^2G(p^2)}\chi\, ,
\end{eqnarray}
from which we can extract the two point function
\begin{eqnarray}
\left<\bar{\chi}\chi\right>^{\text{cont}}~=~i\sigma^\mu p_\mu G(p^2)\, .
\end{eqnarray}
Clearly, in the limit $G(p^2)\rightarrow p^{-2}$, the continuum fermion just reduces to the massless particle limit.
In general, $G(p^2)$ is a complex function whose poles correspond to massive particles and whose branch cut corresponds to a continuum. 
This complex structure is easily captured by introducing the (real-valued) spectral density function $\rho(s)$, such that  
\begin{eqnarray}
G(p^2)~=~\int_0^\infty\,\frac{\rho(s)}{s-p^2-i\e}\,ds~~,~~\rho(s)=\frac{1}{\pi}\text{Im}G(s)\, .
\end{eqnarray}

\begin{figure}
\begin{center}
\includegraphics[width=0.8\textwidth]{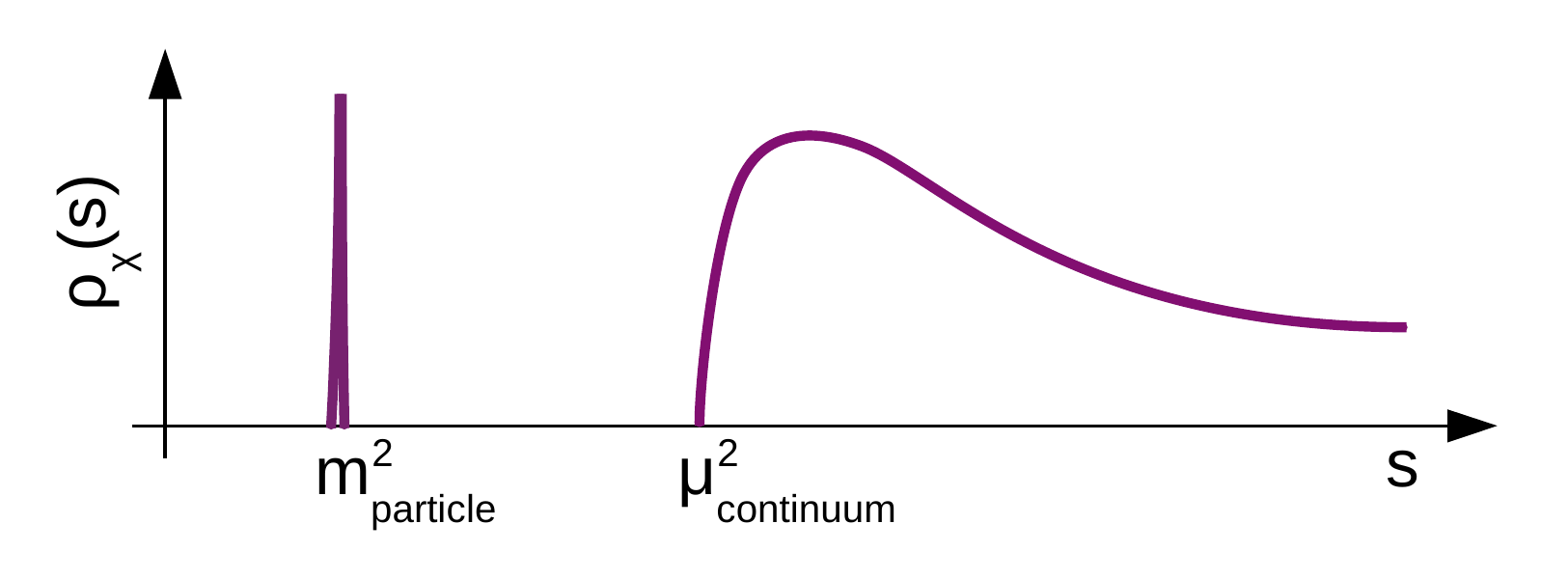}
\end{center}
\caption{A cartoon of a typical fermionic spectral density. The delta function corresponds to a massive particle in the spectrum, while the continuous part indicates a fermion continuum.}
\label{fig:speccartoona}
\end{figure}

\begin{figure}
\begin{center}
\includegraphics[width=0.8\textwidth]{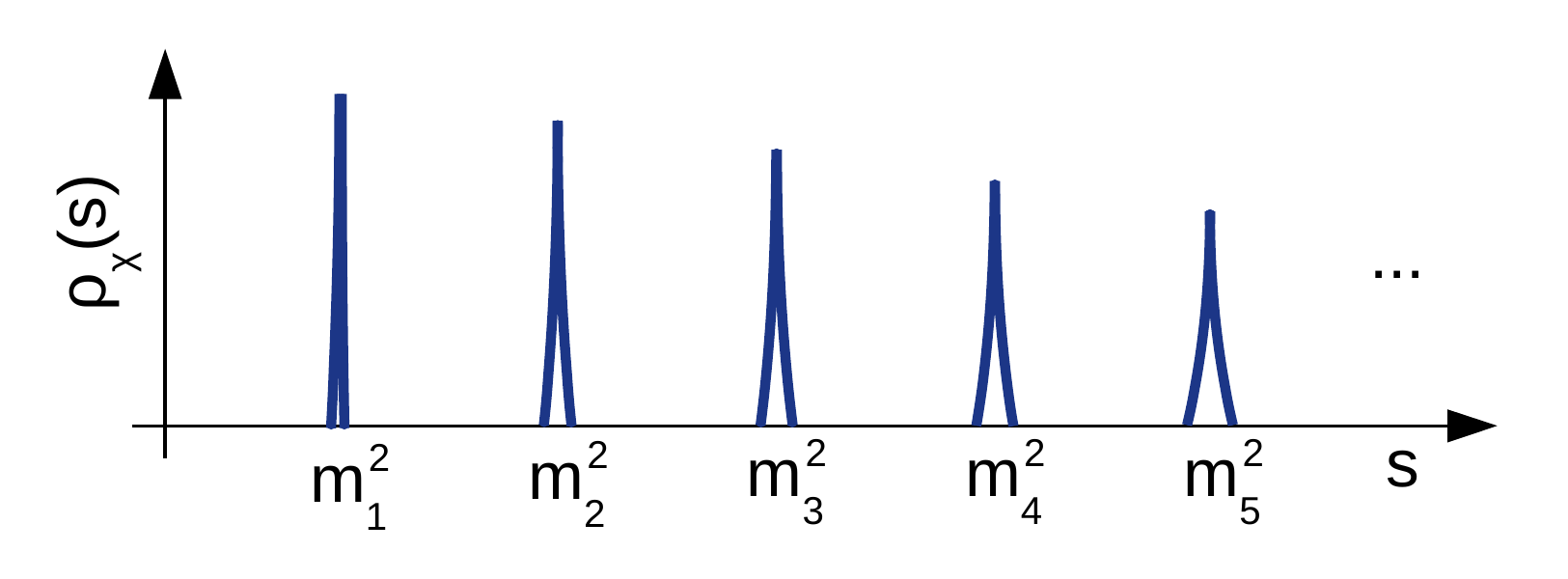}
\end{center}
\caption{A cartoon of a typical fermionic spectral density in the case of an infinite tower of massive fermions (KK modes).}
\label{fig:speccartoonb}
\end{figure}

The spectral density contains all the relevant spectral information for the fermion $\chi$, and is essentially the famous K\"all\'en-Lehmann spectral density \cite{Kallen:1952zz,Lehmann:1954xi}.
Its typical form is illustrated in Fig.~\ref{fig:speccartoona}, with the delta functions corresponding to massive particles and the continuous part encoding the fermion continuum. 
For comparison, in Fig.~\ref{fig:speccartoonb}, we show the spectral density for a tower of massive fermions, which, in the narrow width approximation, is just a sequence of delta functions. 
This is the typical KK spectrum obtained by putting a Weyl fermion in the bulk of 5D Randall-Sundrum geometry~\cite{Randall:1999ee}.
One can indeed think of the continuum as the merging of the spectral density of the KK modes as their separation goes to zero while their width remains finite.
 
\section{Modeling the Continuum Dynamics with Linear Dilaton Geometry} \label{sec:LD}

The effective action presented in the previous section was completely generic in the sense that it did not assume a specific functional form for the spectral density $\rho(s)$. 
However, to say something meaningful about the continuum dynamics, we would like to find a model of the strong dynamics responsible for the emergence of continuum modes that allows us to calculate quantities below the strong scale. 
Inspired by the AdS/CFT duality, we seek to model the continuum dynamics in some weakly coupled, warped 5D geometry. 
We build on past work on how to model continuum dynamics. 
The authors of \cite{Cacciapaglia:2008ns} showed, among other things, how a bulk Dirac fermion in AdS$_5$ is dual to a gapless Weyl fermion continuum, while in \cite{Cai:2009ax}, a gapped supersymmetic continuum arose from a chiral superfield in AdS$_5$ with a bulk dependent mass. 
We will use a setup similar to the latter, albeit in a non-supersymmetric setting.

\begin{figure}
\begin{center}
\includegraphics[width=0.8\textwidth]{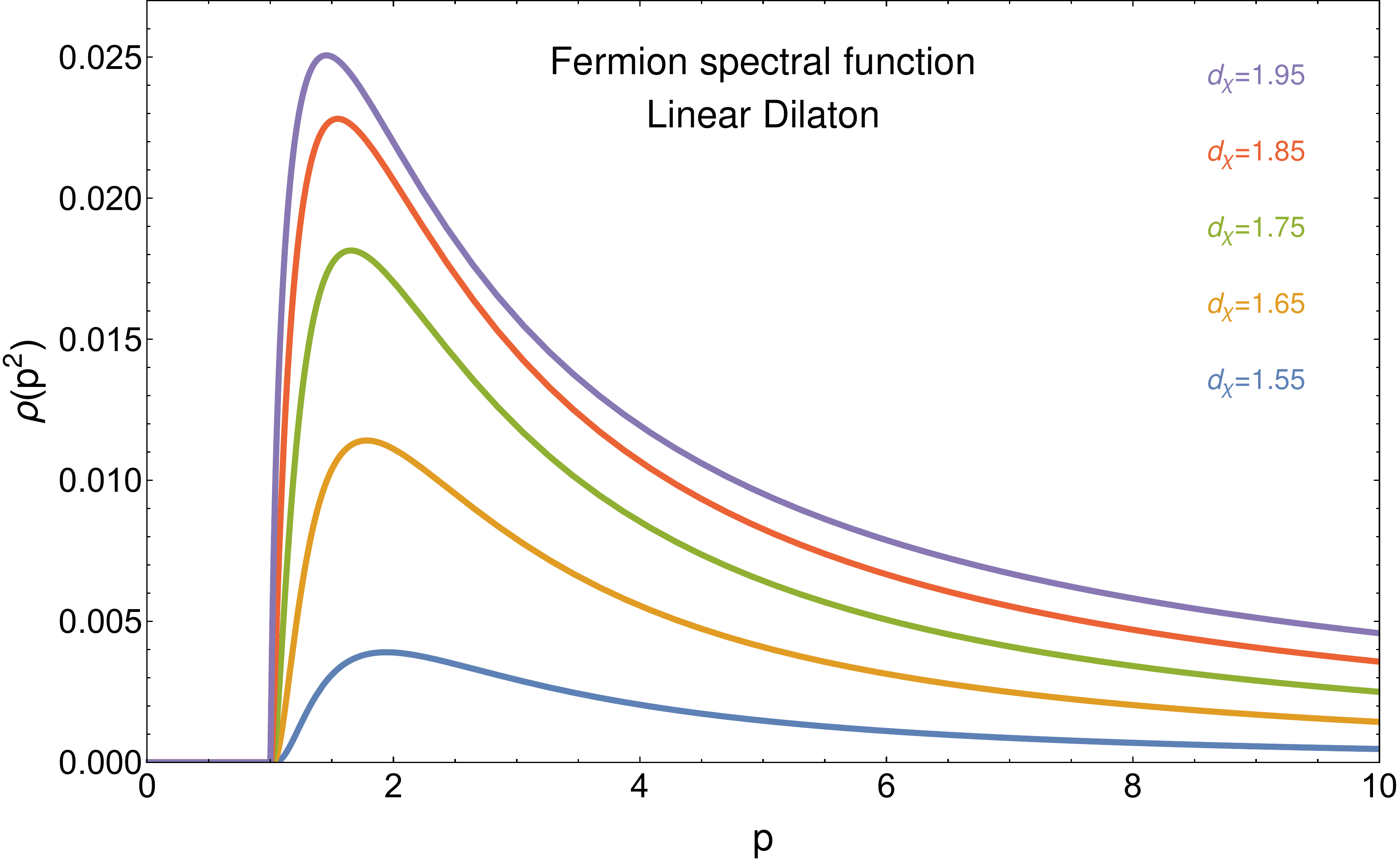}
\end{center}
\caption{The spectral density for a continuum left-handed Weyl fermion $\c$ modeled in a linear dilaton background. The quasi-anomalous dimension $d_\chi$ is linked to the bulk mass by the relation $d_\c=c+2$.}
\label{fig:weylspec}
\end{figure}

To correctly model the continuum dynamics, we consider Weyl fermions in a dilaton background%
\footnote{For the stabilization of linear dilaton backgrounds, see \cite{Giudice:2017fmj} and references therein. 
As we will elaborate below, our realistic model involves an IR brane stabilized by the usual Goldberger-Wise\cite{Goldberger:1999uk} mechanism, which in turn can set the boundary conditions for the linear dilaton, generating the IR scale $\mu$.}. 
In this background the 5D Lagrangian in the string frame is then
\begin{eqnarray}
\mathcal{L}_{S}=e^{-2 \Phi(z)}a^5_S(z)\left[a^{-1}_S(z)\mathcal{L}_{\text{kin}}+\frac{1}{R}\left(c+y\, \Phi(z)\color{black}\right)\left(\psi\chi+\bar{\chi}\bar{\psi}\right)\right]\, ,
\end{eqnarray}
where $z\in\left[0,\infty\right)$ is the coordinate of the fifth dimension, $a_S(z)=\frac{R}{z}$ is the AdS scale factor, $\Phi(z)$ is the dilaton profile, and $y$ is a bulk Yukawa coupling between the dilaton and the bulk fermion. 
Later we will introduce a UV brane and cut off the space at $z=R$. 
The kinetic term is the standard kinetic term for a 5D Dirac fermion:
\begin{eqnarray} \label{eqn:Lagkin}
\mathcal{L}_{\text{kin}}~=~-i\bar{\chi}\bar{\sigma}^\mu p_\mu \chi \,-i \psi\sigma^\mu p_\mu \bar{\psi}\,+\,\frac{1}{2}\left(\psi\overleftrightarrow{\partial}_5 \chi - \bar{\chi}\overleftrightarrow{\partial}_5 \bar{\psi}\right)\, . 
\end{eqnarray}
To conveniently extract the fermion EOM, we first move to the Einstein frame through the rescaling of the coordinates leading to $a(z)=a_S(z)\,e^{-\frac{2}{3}\Phi(z)}$, followed by a canonical renormalization of the fermions. 
The resulting Einstein frame Lagrangian is then
\begin{eqnarray}\label{eq:Lagferm}
\mathcal{L}_{E}=a^4(z)\mathcal{L}_{\text{kin}}+a^5(z)\frac{\hat{c}(z)}{R}\left(\psi\chi+\bar{\chi}\bar{\psi}\right)\, ,
\end{eqnarray}
 where $\hat{c}(z)\equiv(c+y\Phi(z))e^{\frac{2}{3}\Phi(z)}$.
The fermion bulk EOM's are conveniently presented in a Schr\"odinger form \cite{Falkowski:2008fz}:
\begin{eqnarray}
 -\hat{\chi}''(z)+V_{\text{eff}}(z)\,\hat{\chi}(z)~=~p^2\hat{\chi}(z)\, ,
\end{eqnarray}
where $\hat{\chi}(z)={\left(\frac{R}{z}\right)}^2\chi(z)$ and the effective Schr\"odinger potential is
\begin{eqnarray}
V_{\text{eff}}(z)=\frac{c (c+1)+y \Phi(z) (2c+y\Phi(z)+1)-y z \Phi'(z)}{z^2}\, .
\end{eqnarray}
This equation has gapped continuum solutions (similar to scattering solutions in standard QM) when $V_{\text{eff}}(z\rightarrow \infty)=\text{const}>0$. 
That clearly indicates that $\Phi(z)$ has to be linear in $z$ in the deep IR---a linear dilaton. 
For the linear dilaton $\Phi(z)=\mu(z-R)$ with $\mu\sim1$~TeV, $V_{\text{eff}}(z\rightarrow \infty)=y^2\mu^2$, and we expect a continuum beyond the gap $y\mu$. 
Indeed, the IR regular%
\footnote{As in standard AdS/CFT, we define the "IR regular" solution for Lorentzian AdS as the analytic continuation of the corresponding IR regular solution for Euclidean AdS. 
This is equivalent to choosing an outgoing wave boundary condition in Lorentzian AdS.} 
bulk solutions are
\begin{equation} \begin{split}
\chi(z)\,&=\,A\,a^{-2}(z)\,\,W\left(-\frac{c\mu y}{\Delta},c+\frac{1}{2},2\Delta z\right) \,, \\
\psi(z)&=A\,a^{-2}(z)\,\,W\left(-\frac{c\mu y}{\Delta},c-\frac{1}{2},2\Delta z\right)\frac{\mu y-\Delta}{p}\, ,
\end{split} \end{equation}
where $\Delta=\sqrt{y^2\mu^2-p^2}$ and $W(a,b,z)$ is a Whittaker function. 
From these bulk solutions we can extract the left-handed (LH) source Green's function as
\begin{eqnarray}
\left\langle\mathcal{O}_{\text{R}}\mathcal{O}_{\text{R}}\right\rangle~=~-\frac{{(2\pi)}^{-2c-1} \left(\mu  y-\Delta\right)}{2\left(c+\frac{1}{2}\right) p^2 }\,\frac{\Gamma (1-2 c) }{ \Gamma (1+2 c)}\,\frac{\Gamma \left(1+c\,\frac{\mu  y+\Delta}{\Delta}\right)}{\Gamma \left(1+c\frac{\mu  y-\Delta}{\Delta}\right)}\, {(2\Delta)}^{2c} \, ,
\end{eqnarray}
while the Green's function for $\chi$ is its (almost) inverse
\begin{eqnarray}
G(p^2)~=~-\frac{{(2\pi)}^{2c-1}}{2\left(c-\frac{1}{2}\right) \left(\mu  y-\Delta\right)}\,\frac{\Gamma (1+2 c) }{ \Gamma (1-2 c)}\,\frac{\Gamma \left(1+c\,\frac{\mu  y-\Delta}{\Delta}\right)}{\Gamma \left(1+c\frac{\mu  y +\Delta}{\Delta}\right)}\, {(2\Delta)}^{-2c}\, .
\end{eqnarray}
The notation $\overline{\lim}$ indicates a regulated limit, i.e., the leading term regulated by powers of $z$. 
For a right-handed (RH) source, we send $\chi(z) \leftrightarrow \psi(z)$ and $y\rightarrow -y$. 
The Green's function, extracted from 5D, now serves as the momentum-dependent form factor of Eq.~\ref{eq:WeylCont}.
It has a pole at $p^2=0$, indicating a massless zero mode.
For $p\geq y\mu$, $\Delta$ goes imaginary and $G(p^2)$ has a branch cut corresponding to the continuum. 
The exact form of the spectral density depends on the bulk mass $c$, which we take in the range $0\leq c<\frac{1}{2}$ to avoid poles from the gamma functions. 
The resulting spectral densities for select values in this range are shown in Fig.~\ref{fig:weylspec}. 
Note also that following the case of pure AdS$_5$ \cite{Cacciapaglia:2008ns}, we can assign the LH source an anomalous dimension $d_{\mathcal{O}_{\text{R}}}=2+c$ in the range $2\leq d_{\mathcal{O}_{\text{R}}}\leq \frac{5}{2}$
\footnote{For a RH source, the identification becomes $d_{\mathcal{O}_{\text{L}}}=2-c$ and $3/2 \leq d_{\mathcal{O}_{\text{R}}}\leq 2$.}. This anomalous dimension governs the UV behavior of our spectral function, where it goes over to the fermionic unparticle spectral function. The corresponding anomalous dimension for the LH fermion is given by $d_{\mathcal{\chi}}=4-d_{\mathcal{O}_{\text{R}}}=2-c$.

\begin{figure}
\begin{center}
\includegraphics[width=0.8\textwidth]{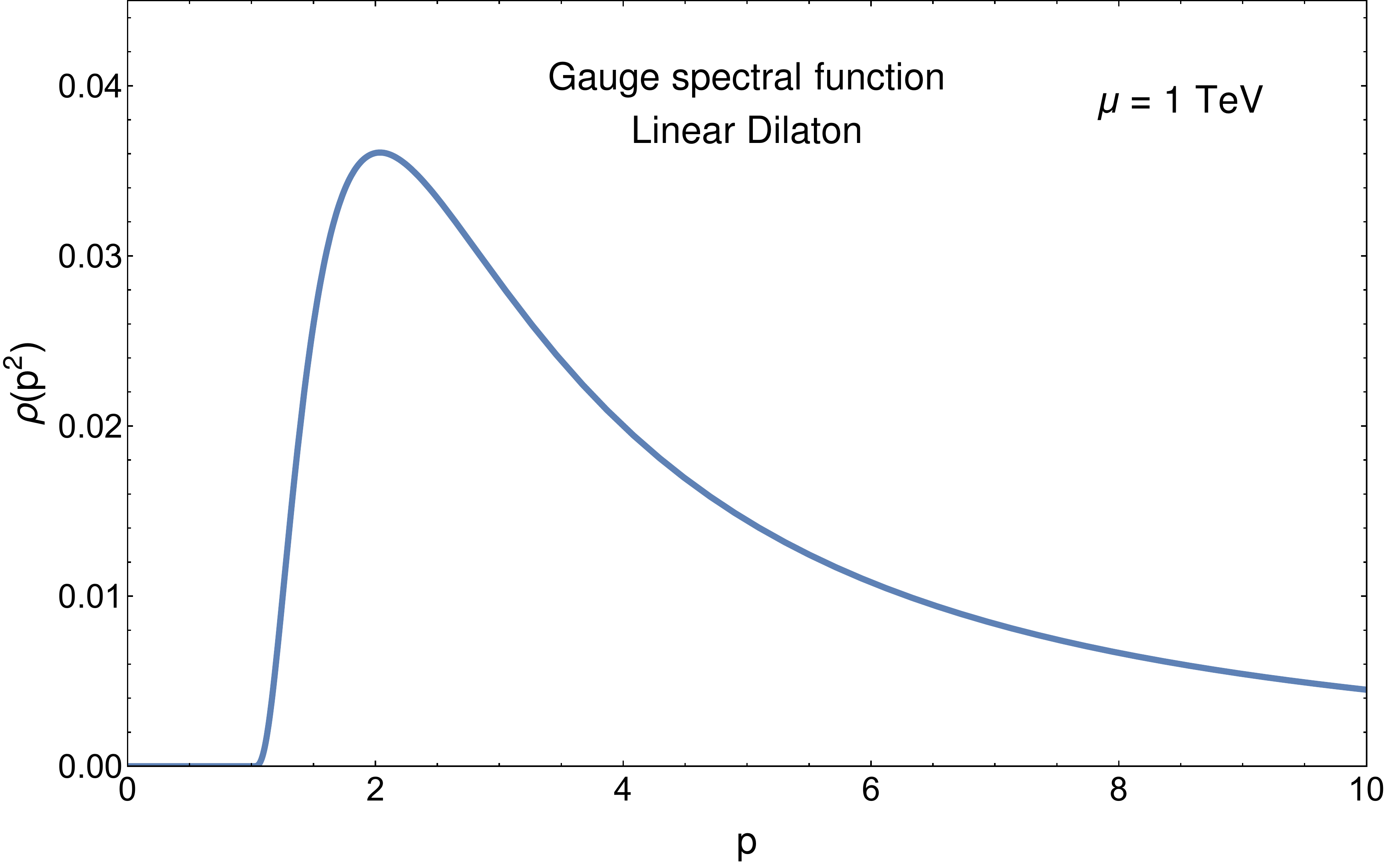}
\end{center}
\caption{The spectral density for a continuum gauge boson in a linear dilaton background.}
\label{fig:gaugespec}
\end{figure}

We model the gauge continuum in a similar way to the fermion continuum, by considering gauge modes in the bulk of a linear dilaton geometry with $\Phi(z)=\mu (z-R)$. 
The Einstein frame Lagrangian is
\begin{eqnarray}
\mathcal{L}_{\text{E}}=a(z)\,e^{-\frac{4}{3}\mu (z-R)}\left[\frac{1}{4}F^{MN}F_{MN}\right]\, ,
 \end{eqnarray}
while the effective Schr\"odinger equation is
\begin{eqnarray}
 -\hat{A}''(z)+V_{\text{eff}}(z)\hat{A}(z)~=~p^2\hat{A}(z)\, ,
\end{eqnarray}
 where $\hat{A}(z)=\sqrt{\frac{R}{z}}\,e^{-\mu(z-R)}\,A(z)$ and the effective Schr\"odinger potential is
\begin{eqnarray}
V_{\text{eff}}(z)=\mu ^2+\frac{\mu }{z}+\frac{3}{4 z^2} \, .
\end{eqnarray}
As in the fermion case, the potential in the deep IR goes to a constant, $V_{\text{eff}}(z\rightarrow \infty)=\mu^2$. 
Hence we expect a gauge continuum with a gap of $\mu$. 
Indeed, the IR regular bulk solutions are
\begin{eqnarray}
A(z)~=~A\,\sqrt{\frac{z}{R}}\, e^{\mu  (z-R)} \,W\left(-\frac{\mu}{2\Delta},1;2\Delta z\right)\,,
\end{eqnarray}
with $\Delta=\sqrt{\mu ^2-p^2}$, and the Neumann Green's function has a pole at $p^2=0$ and a branch cut for $p^2>\mu^2$ with the spectral density
\begin{eqnarray}
\rho(s)~=~\frac{1}{\pi} \underset{z\rightarrow 0}{\overline{\lim}} \,\text{Im}\frac{A(z)}{A'(z)}=
\frac{1}{2\pi s}\left[1+ i\psi\left(\frac{1}{2}+\frac{\mu }{2 \Delta}\right)-i\psi\left(\frac{1}{2}-\frac{\mu
   }{2 \Delta}\right)\right]\, ,
\end{eqnarray}
where $\psi(x)$ is the digamma function. 
The gauge spectral density is depicted in Fig.~\ref{fig:gaugespec}.

\section{A Realistic Continuum Composite Higgs Model}\label{sec:fullmod}

In the two previous sections, we have shown how to model the fermion and gauge continua in a linear dilaton geometry. 
Here, we use them as building blocks in a full CH model, in which the continuum fermion and gauge modes play the role of top and gauge partners. 
In fact, our construction mirrors the existing CH models in its group theory, choice of representations, etc. 
The only modification is in the introduction of a linear dilaton geometry instead of the standard RS one.

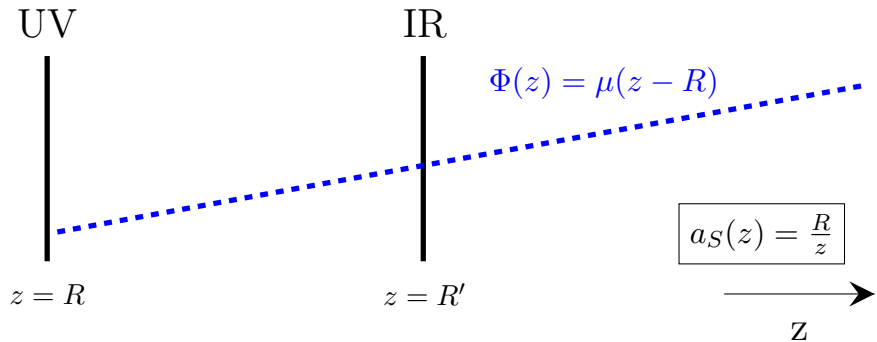
\begin{figure}
\begin{center}
\begin{tikzpicture}[baseline=(a)]
\begin{feynman}[]
\vertex (a){};
\vertex[below=3cm of a] (b){};
\vertex[right=5cm of a] (c){};
\vertex[below=3cm of c] (d){};
\vertex[above=0.5cm of b] (e){};
\vertex[right=5cm of e] (f){};
\vertex[right=6cm of f] (g){};
\vertex[above=2cm of g] (h){};
\diagram* {
 (a) --[line width=0.7mm] (b),
 (c) --[line width=0.7mm] (d),
 (e) --[line width=0.7mm,dashed,blue] (h)
};
\end{feynman}
\draw [decoration={markings,mark=at position 1 with
    {\arrow[scale=3,>=stealth]{>}}},postaction={decorate}] (9,-3.3) -- (11,-3.3);
\node[scale=1.4] at (10,-3.8) {z};
\node[scale=1.4] at (0,0.3) (aa) {UV};
\node[scale=1] at (0,-3.3) {$z=R$};
\node[scale=1.4] at (5.02,0.3) () {IR};
\node[scale=1] at (5.02,-3.3) {$z=R'$};
\node[scale=1.1,blue] at (7.4,-0.5) {$\Phi(z)=\mu(z-R)$};
\node[scale=1.1] at (9.5,-2.5) () {$a_S(z)=\frac{R}{z}$};
\draw [] (8.4,-2.9) rectangle (10.6,-2.1);
\end{tikzpicture}
\end{center}
\caption{A sketch of our geometry in the string frame. The IR brane carries local fields that result in jump conditions for the bulk fields.}\label{fig:5D}
\end{figure}

Our extra dimensional geometry is depicted in Fig.~\ref{fig:5D}. We consider AdS$_5$ regulated by a UV brane at $z=R$. 
In addition we introduce an IR brane at $z=R'\sim 1/\text{TeV}$, which is stabilized as usual by the Goldberger-Wise mechanism. 
The fifth dimension continues beyond the IR brane to $z\rightarrow\infty$. 
The IR brane has a double role in our model: 
\begin{enumerate}
\item It provides the location for the breaking of the bulk gauge symmetry.
\item It is responsible for the generation of the IR scale $\mu\sim\text{TeV}$, which is the slope of the linear dilaton. 
The dilaton profile is basically negligible up to distances close to the IR brane, where it has a boundary condition involving the IR scale $\mu$. 
After the IR brane, the dilaton grows linearly with a slope $\mu$. 
We are agnostic about the exact mechanism stabilizing the dilaton beyond the IR brane (see \cite{Giudice:2017fmj} and references therein for possibilities), but note that there is no tuning involved because the slope of the dilaton is related to its boundary condition on the IR brane. 
In other words, the solution to the hierarchy problem in our case is the usual Goldberger-Wise mechanism, or dimensional transmutation. 
The linear dilaton is merely a way to model a different confining dynamics which gives rise to composite continua. 
\end{enumerate}

The remaining details of our model are very similar to standard CH models \cite{Contino:2003ve,Agashe:2004rs,Medina:2007hz,Csaki:2008zd}. 
We consider a $G = SO(5)\times U(1)_X$ gauge symmetry in the bulk of our geometry. 
This gauge symmetry is reduced to $SO(4)\times U(1)_X$ on the IR brane, by giving Dirichlet boundary conditions (B.C.) to the gauge fields corresponding to broken generators. 
On the UV brane, we break $SO(5)\times U(1)_X$ to the SM electroweak gauge symmetry $SU(2)_L\times U(1)_Y$, such that $Y=T^3_R+X$. 
This choice of boundary conditions leads to a zero mode in the fifth component of the bulk gauge field, $A^{\hat{a}}_5$, with $\hat{a}$ denoting the generators in the coset $G/H$. 
The role of the pNGB Higgs boson is then played by the Wilson line from the UV to the IR brane, $ig_5\int_R^{R'}\,A_5\,dz$. 
It is interesting to note that this Wilson line between the two branes is the only gauge invariant Wilson line we can write, so there is no physical meaning to the $A_5$ profile beyond the IR brane. 
We can always account for the effect of the $A_5$ vacuum expectation value (VEV) by rotating it into the matching conditions on the IR brane. 

In addition to the bulk gauge symmetry, we embed the SM fermions $q_L,\,t_R,\,b_R$ in the bulk multiplets $Q_L,\,T_R,\,B_R$, transforming in the $\mathbf{5}_{\frac{2}{3}},\, \mathbf{5}_{\frac{2}{3}},\,\mathbf{10}_{\frac{2}{3}}$ representations of $SO(5)\times U(1)_X$, respectively. 
This is the same choice of bulk representations as \cite{Medina:2007hz,Csaki:2008zd}. 
Under the subgroup $SU(2)_L\times U(1)_Y$, the bulk multiplets decompose as:
\begin{eqnarray}\label{eq:decomp}
Q_L(\mathbf{5})_{\frac{2}{3}}~&\rightarrow&~q_L(\mathbf{2})_{\frac{1}{6}}~+~\tilde{q}_L(\mathbf{2})_{\frac{7}{6}}~+~y_L(\mathbf{1})_{\frac{2}{3}}\,,\nonumber\\
T_R(\mathbf{5})_{\frac{2}{3}}~&\rightarrow&~q_R(\mathbf{2})_{\frac{1}{6}}~+~\tilde{q}_R(\mathbf{2})_{\frac{7}{6}}~+~t_R(\mathbf{1})_{\frac{2}{3}}\,, \\
B_R(\mathbf{10})_{\frac{2}{3}}~&\rightarrow&~q'_R(\mathbf{2})_{\frac{1}{6}}~+~\tilde{q}'_R(\mathbf{2})_{\frac{7}{6}}~+~x_R(\mathbf{3})_{\frac{2}{3}}~+~y_R(\mathbf{1})_{\frac{7}{6}}~+~\tilde{y}_R(\mathbf{1})_{\frac{1}{6}}~+~b_R(\mathbf{1})_{-\frac{1}{3}}\, . \nonumber
\end{eqnarray}
Let $\chi$ and $\psi$ be the LH and RH components of the bulk Dirac fermion appearing in \eqnref{Lagkin}.
On the UV brane, the states $\chi_{q_L},\, \psi_{t_R}$ and $\psi_{b_R}$ get Neumann B.C., while all other states in $\chi_{Q_L},\, \psi_{T_R}$ and $\psi_{B_R}$ get Dirichlet B.C. On the IR brane, all of the states in $\chi_{Q_L},\, \psi_{T_R}$ and $\psi_{B_R}$ get Neumann B.C. 
Consequently, we have zero modes only for the SM states $q_L,\,t_R$ and $b_R$. 
On the IR brane with induced metric $g_{\text{ind}}$, we can write the $SO(4)\times U(1)_X$ invariant mass terms:
\begin{eqnarray}\label{eq:IRlag}
S_{\text{IR}}~=~\int~d^4x~\sqrt{g_{\text{ind}}}~\left[\,M_1 \, \bar{y}_L t_R~+~M_4 \, \left(\bar{q}_L q_R + \bar{\tilde{q}}_L \tilde{q}_R\right)~+~M_b\,  \left(\bar{q}_L q'_R + \bar{\tilde{q}}_L \tilde{q}'_R\right)\,\right]\,.
\end{eqnarray}
These terms give rise to the SM Yukawa coupling in the 4D effective action. 
From the 5D point of view, these IR brane-localized terms provide the discontinuity (jump B.C.) resulting in quasi-IR brane-localized wave function profiles for the fermionic fields (albeit with support in the deep IR), but with large enough wave function overlap with the physical $A_5$ below the IR brane to obtain the correct top mass.

\section{Summary of Results\label{sec:summary}}

In this section we present a concise summary of the results in our model. The details are fleshed out in the next sections and the appendices. 

\begin{itemize}
\item \textit{Overview}: 
We constructed a realistic CH model with continuum top partners. 
There are no fermionic KK resonances in the model. The continuum generically does contain broad peaks (of width $\sim\text{TeV}$) that could be probed with non-resonant high $p_T$ dilepton searches at a future $100$~TeV collider.
The only gauge particle resonances occur at energies well outside the reach of the LHC. 

\item \textit{Model parameters}:
We have only two additional parameters to the standard parameters of CH models: the dilaton slope $\mu$ and the fermion-dilaton Yukawa $y$. 
The other standard CH parameters are $R$ and $R'$, as well as the gauge parameters $\theta,\,r$ (see \secref{specden}) and the fermion bulk and IR brane mass parameters $c_Q, c_T, c_B, M_1, M_4$ and $M_b$. 
We demonstrate a realistic SM spectrum and Higgs potential for the following benchmark point (BP) in parameter space:  
\begin{equation} \begin{split} \label{eq:benchmark}
R/R'=10^{-16}, \ 1/R'=2.81 & \text{ TeV}, \ \mu=1 \text{ TeV}, \ y=1.75, \\
r=0.975, & \ \sin\theta=0.39, \\
c_Q=0.2, \ c_T=&-0.22, \ c_B=-0.03, \\
M_1=1.2, \ M_4 &=0, \ M_b=0.017 \,.
\end{split} \end{equation}

For this particular point in parameter space, all of the SM variables are correctly reproduced, except for the mass of the top quark, which is a bit too light ($125$~GeV instead of $140$~GeV at $2$~TeV). 
This is an artifact of our particular bulk fermion representations that also exists in standard CH models \cite{Medina:2007hz,Csaki:2008zd}, and can be overcome by either changing to different bulk representations or choosing a slightly more tuned point in parameter space.

\item \textit{Continua:} 
In our specific point in parameter space, the gauge continuum starts at $\mu=1$~TeV, while the fermion continuum starts at $y\mu=1.75$~TeV.
It is of course possible to find other points with larger gaps for the continua, at the cost of more tuning in the Higgs potential. 
The existence of the relatively low gauge continuum is phenomenologically viable due to the lack of $s$-channel resonances in our model, but the fermion continuum taken to be higher to avoid tension with LHC bounds. 
The spectral densities of the top and bottom are depicted in Fig.~\ref{fig:fullf} and those of the $W$ and $Z$ in Fig.~\ref{fig:fullg}. 
The broad peaks in the fermion spectral densities at $5$~TeV and $9$~TeV originate in their IR brane masses (see Sec.~\ref{sed:fermspec}). These broad peaks, of width $\sim1$~TeV, could be probed at a future $100$~TeV collider.

\begin{figure}
\begin{center}
\includegraphics[width=0.6\textwidth]{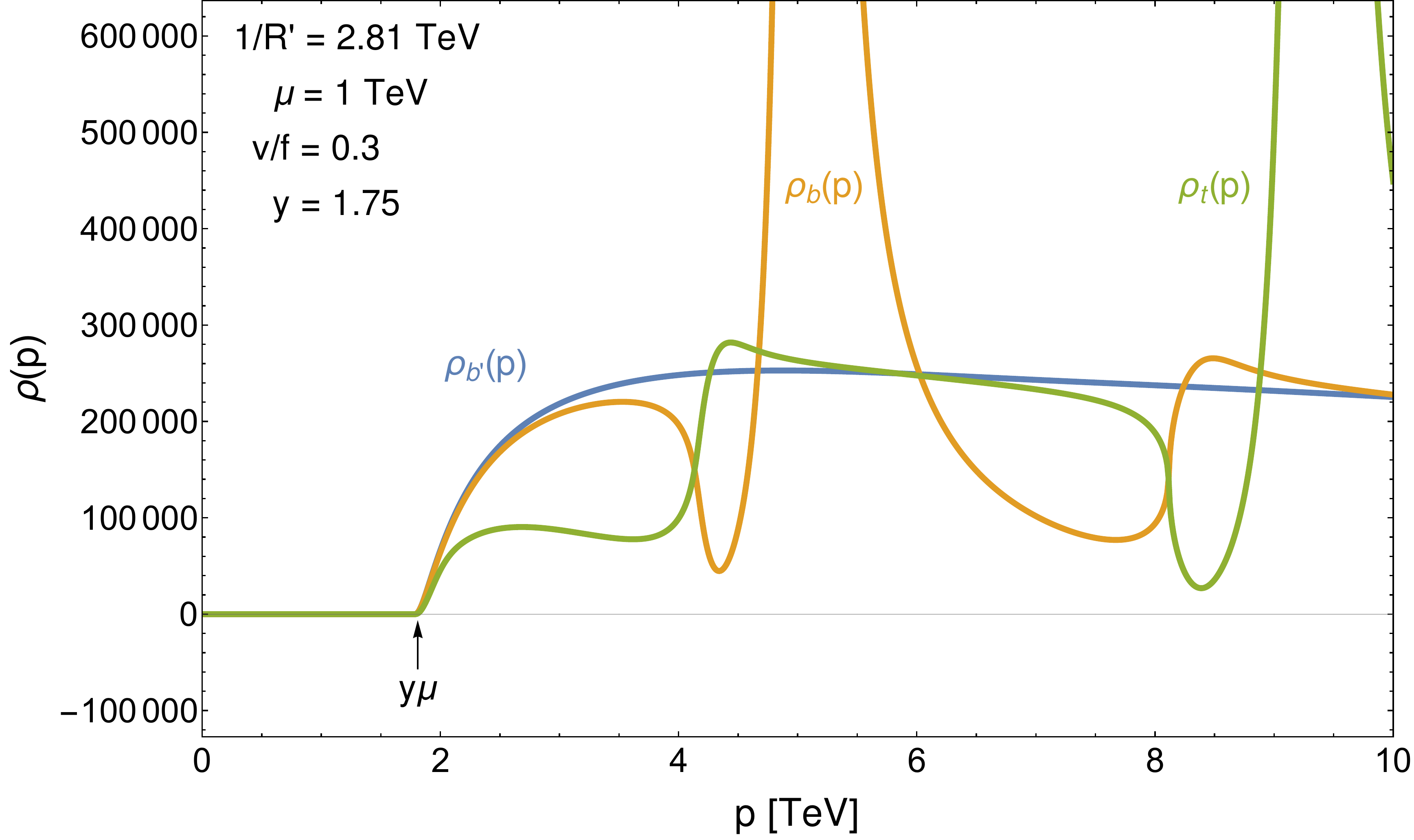}
\end{center}
\caption{Top, bottom and $b^\prime$ spectral densities for $v/f=0.3$ and parameter values from the BP in Eq.~\ref{eq:benchmark}. The spectral function features broad peaks that could be probed at a future $100$~TeV collider.} 
\label{fig:fullf}
\end{figure}

\begin{figure}
\begin{center}
\includegraphics[width=0.6\textwidth]{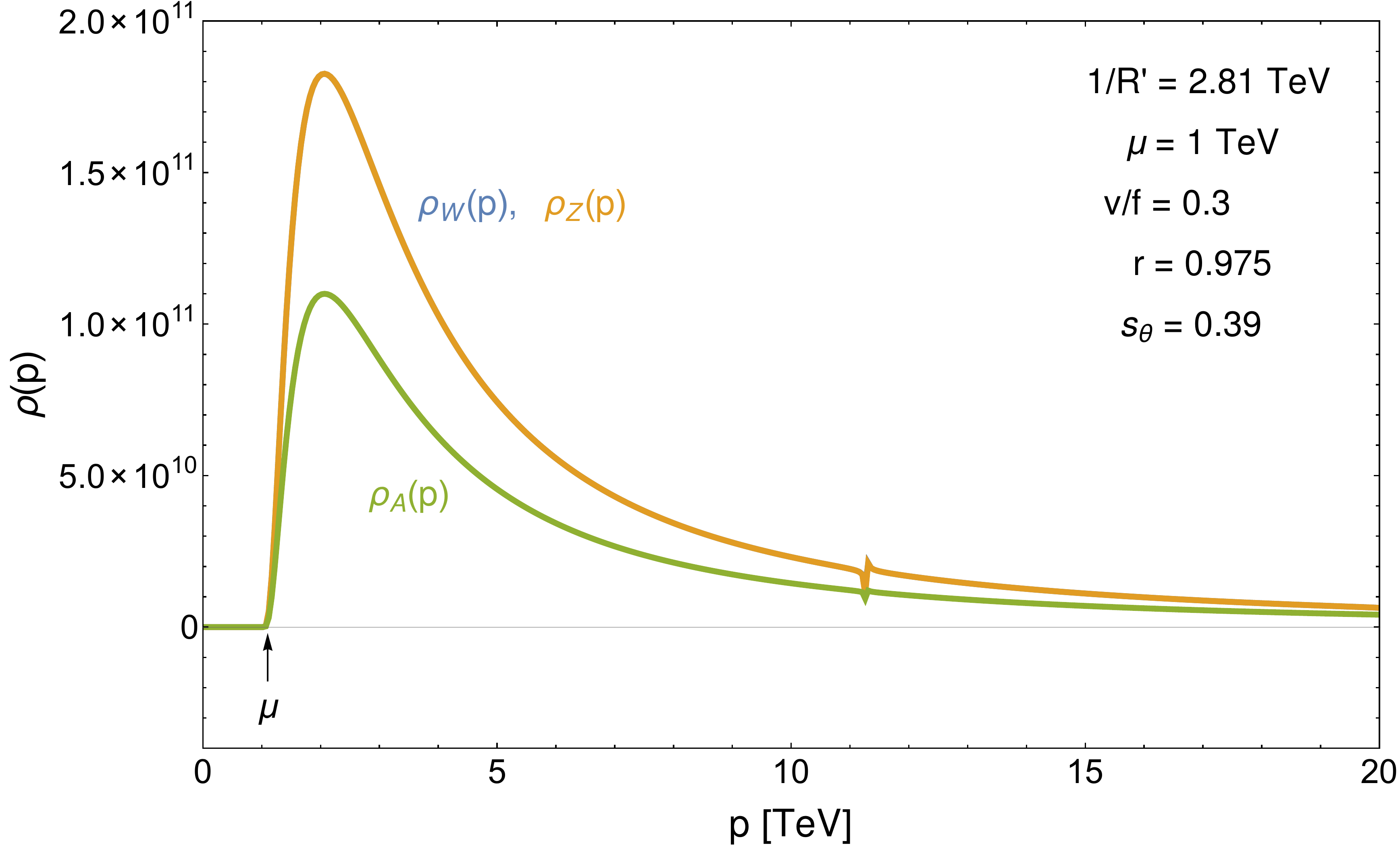}
\end{center}
\caption{Gauge spectral density for $v/f=0.3$ and parameter values from the BP in Eq.~\ref{eq:benchmark}.}
\label{fig:fullg}
\end{figure}

\item \textit{Higgs Potential:} 
As in standard CH models, the potential for the pNGB Higgs is radiatively generated. 
The radiative contributions of the top, $W$ and $Z$ in our model are balanced by the contribution of the fermion and gauge continua, which also couple to the pNGB Higgs. 
We get the correct Higgs potential at the cost of a standard 1\% tuning, with $v/f=0.3$, consistent with electroweak precision bounds \cite{Agashe:2005dk}. This is to be compared with per mille level tuning in a corresponding composite Higgs model with the same IR scale $R'$ and the same choice of bulk representations \cite{Panico:2012uw}. 
The Higgs potential in our model is depicted in Fig.~\ref{fig:Higgspot}.

\begin{figure}
\begin{center}
\includegraphics[width=0.6\textwidth]{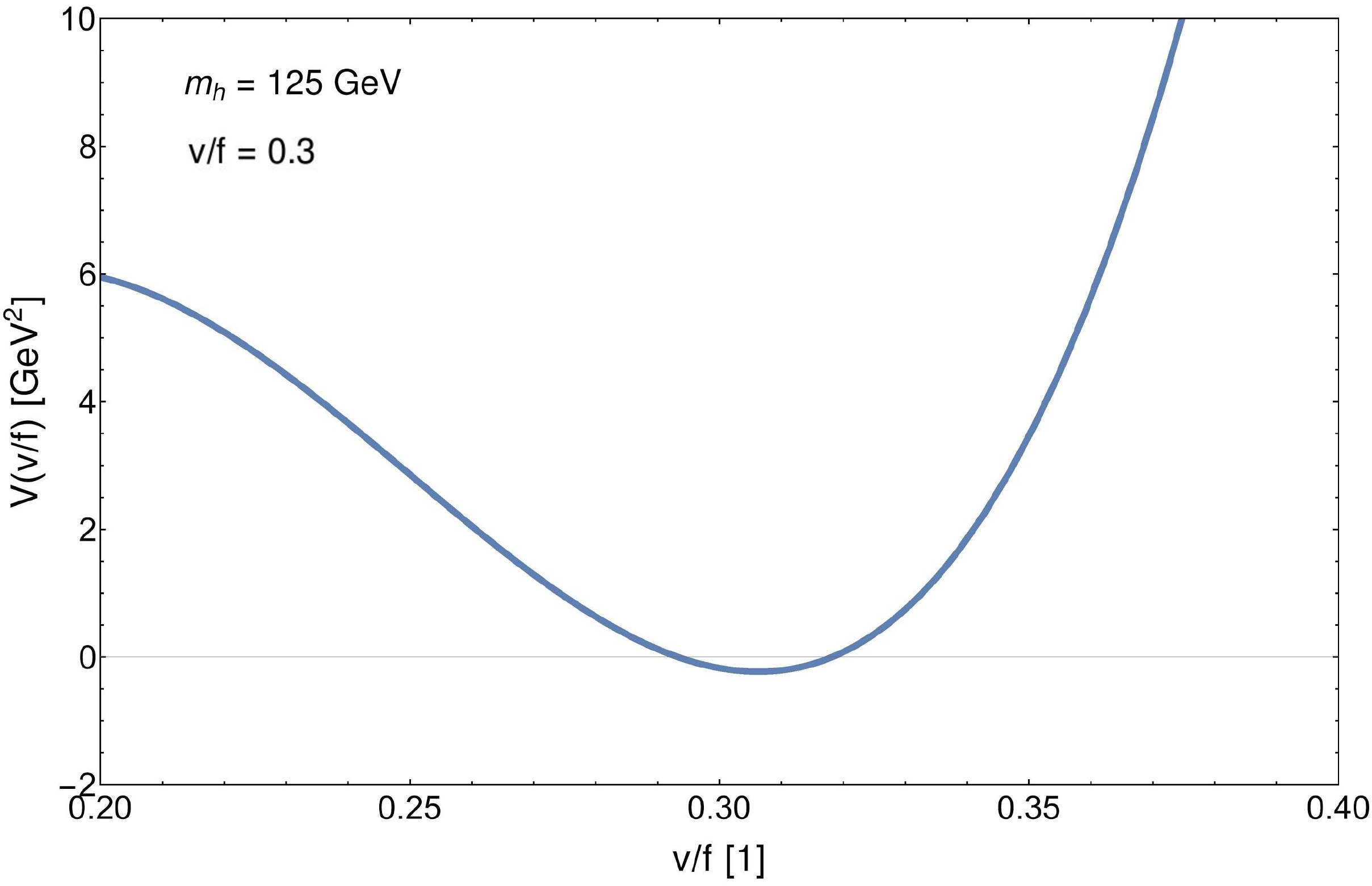}
\end{center}
\caption{The Coleman-Weinberg potential in our model. 
The minimum is at $v/f=\sin\left(\left<h\right>/f\right)=0.3$ and the Higgs mass is reproduced.}
\label{fig:Higgspot}
\end{figure}

\item \textit{Phenomenology:} 
Since there are practically no $s$-channel resonances in our model, the regular bounds on KK gauge bosons do not apply. 
This is the main feature of the continuum naturalness, which is illustrated in Fig.~\ref{fig:schannel}, where the partonic cross-section for $\sigma\left(q\bar{q}\rightarrow G^*\rightarrow t_R\bar{t}_R\right)$ in our model is compared to models with gauge KK resonances. We use $G^*$ to denote the overall sum of the SM gluon, the tower of KK gluons in the RS case, and the gluon spectral density in the continuum case. Compared to the KK case, in the continuum case the spectral density tends to push the effect of new physics to the higher invariant mass regions where there is a larger PDF suppression, resulting in an overall suppression of the total cross section. To simplify the calculation we assume an IR brane localized $t_R$, which is in general a very good approximation in CH models. 

\begin{figure}
\begin{center}
\includegraphics[width=0.6\textwidth]{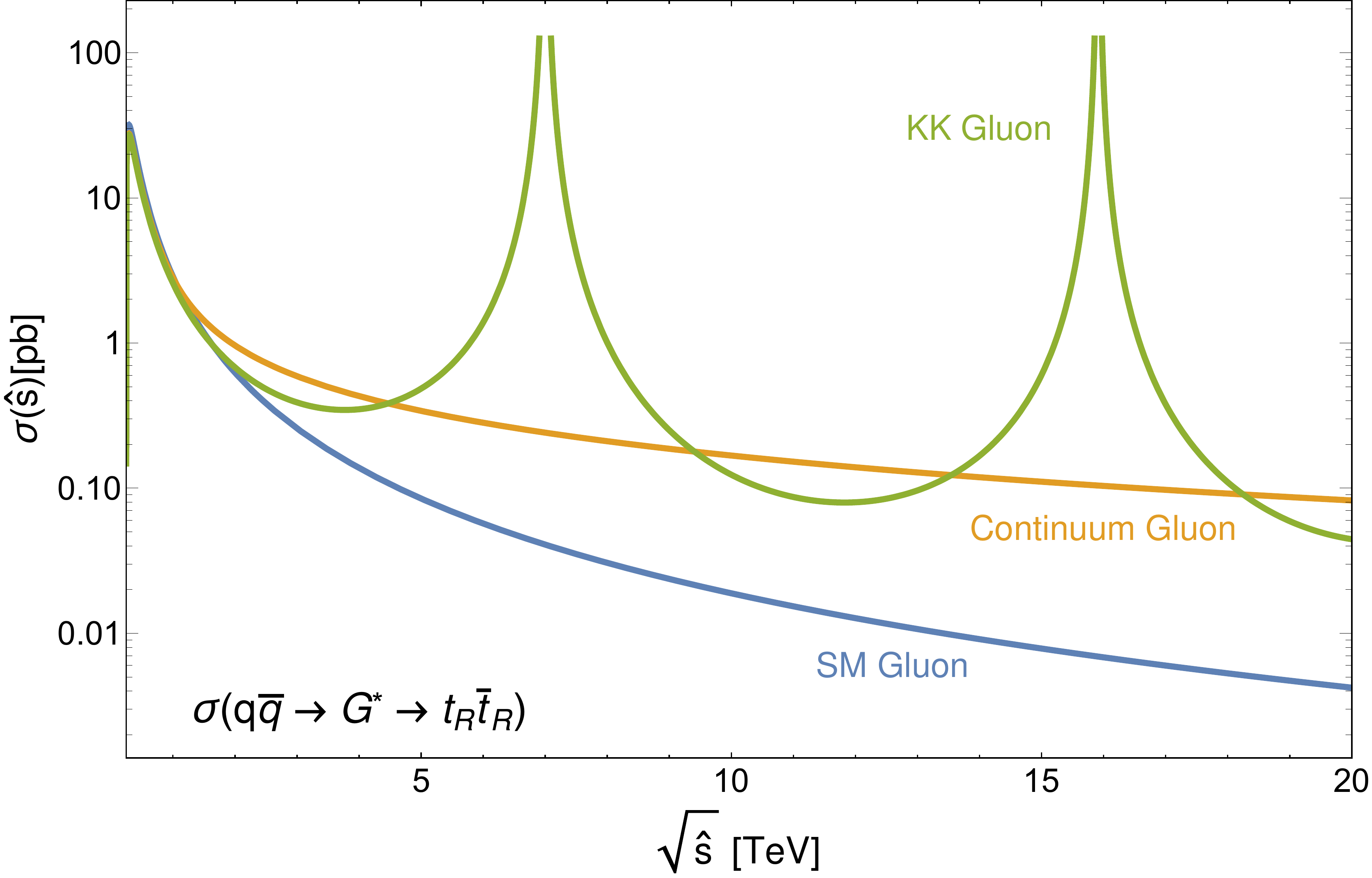}
\end{center}
\caption{The partonic cross section $\sigma\left(q\bar{q}\rightarrow G^* \rightarrow t_R\bar{t}_R\right)$ in three simplified models: only SM gluon, with KK gluons, and with continuum gluons. 
While the presence of KK gluons leads to resonances in the partonic cross section, the continuum only leads to a smooth rise above the SM background.}
\label{fig:schannel}
\end{figure}

A lower bound on the gap of the gauge continuum is obtained from the running of $\alpha_s$ in Sec.~\ref{sec:pheno}: the bound is $\mu \gtrsim 600$~GeV.
To correctly infer the LHC bounds on continuum top partners, we have to calculate their pair production cross-section, which we do in an upcoming work \cite{CLLT}. 

\end{itemize}

\section{Calculating Spectral Densities}~\label{sec:specden}

In the following sections we give an overview of the calculations that lead to the results in Sec.~\ref{sec:summary}. 
Some of the details are presented in the appendices. 
This section explains the basic ingredients used for obtaining the various spectral densities for the gauge and fermion states in our model, using methods similar to Sec.~\ref{sec:LD}. 
The 4D Green's function is extracted as the UV limit of the 5D Green's function as
\begin{eqnarray}\label{UVGreen}
G(p^2)~=~\lim_{z,z'\rightarrow R}\,G(z,z';p^2)\, ,
\end{eqnarray}
where $G(z,z';p^2)$ is the IR regular solution to the inhomogeneous EOM, which is schematically
\begin{eqnarray}\label{eq:Greenschem}
\mathcal{D}_{g/f} G(z,z';p^2)~=~\delta(z-z')\, ,
\end{eqnarray}
subject to the UV B.C. and IR matching conditions. In the above equation, $\mathcal{D}_{g/f}$ is the relevant differential operator for the gauge bosons/fermions.
The spectral densities are obtained from the Green's functions via $\rho(s)=\frac{1}{\pi}\text{Im}\,G(s)$. 

\subsection{Gauge Boson Spectral Densities} \label{sec:gaugespecden}

The homogeneous EOM for gauge fields is 
\begin{eqnarray}\label{eq:gaugeeomd}
a_{\text{eff}}(z)\,p^2 G~+~ \plz \left[a_{\text{eff}}(z)\,\plz G\right]~=0\, .
\end{eqnarray}
where $a_{\text{eff}}(z)\equiv\left(\frac{R}{z}\right)e^{-2\mu(z-R)}$. The general solution to this equation is
\begin{eqnarray}\label{eq:bulksolg}
G(z)~&=&~{\left(\frac{z}{R}\right)}\,e^{\mu(z-R)}\,\left[A\,M\left(-\frac{\mu}{2\Delta},1;2\Delta z\right)\,+\,B\,W\left(-\frac{\mu}{2\Delta},1;2\Delta z\right)\,\right]\, ,
\end{eqnarray}
where $W$ and $M$ are Whittaker functions, and $A$ and $B$ are coefficients.
To find the gauge Green's functions, we have to solve the inhomogeneous version of Eq.~\ref{eq:gaugeeomd}, with $\delta(z-z')$ inserted on the right hand side, subject to the boundary conditions presented below.

\subsubsection{UV Boundary Conditions} 

The UV B.C. are Neumann for $SU(2)_L\times U(1)_Y$ and Dirichlet for the other generators. In particular, we give Neumann B.C. for the hypercharge gauge boson $B_\mu$ defined as:
\begin{eqnarray}
B_\mu = \sin\theta~W^{3R}_\mu + \cos\theta~X_\mu\, .
\end{eqnarray}
where $W^{3R}_\mu$ is the gauge boson corresponding to the $T^3_R$ generator of $SO(5)$ and the $X_\mu$ gauge boson of the $U(1)_X$. 
The angle $\theta$ is set to reproduce the correct Weinberg angle $\theta_W$.
Similarly to \cite{Csaki:2008zd}, we also include a UV brane localized kinetic term for the $SU(2)_L$ gauge bosons, which changes their UV B.C. to 
\begin{eqnarray}
\partial_z G_{SU(2)_L} +  C_{BKT}\,G_{SU(2)_L}\,|_{z=R}=0\,
\end{eqnarray}
where $C_{BKT}=r^2\,p^2 R \log R'/R$ and $r$ is an $O(1)$ parameter adjusted to reproduce the correct right $W$ mass for any value of $g_5$.

\subsubsection{IR Matching Conditions}

On the IR brane we have matching conditions, which are either
\begin{eqnarray}
\Delta G|_{z=R'}~=~\Delta \plz G|_{z=R'}~=~0\, 
\end{eqnarray}
for the generators of $SO(4)\times U(1)_X$ or
\begin{eqnarray}
G|_{z=R'^-}~=~G|_{z=R'^+}~=~0\, 
\end{eqnarray}
for the generators of $SO(5)/SO(4)$. Here and below, we use $G|_{z=R'^+}, G|_{z=R'^-}$ to denote the Green's function evaluated at $z\rightarrow R'$ from the right and left, respectively, and $\Delta G \equiv G|_{z=R'^+} - G|_{z=R'^-}$. We only select Green's functions that are regular in the deep IR, as described in Sec.~\ref{sec:LD}. 

\subsubsection{Accounting for the VEV of the Higgs--$A_5$}

As a final complication, note that we are generically looking for the Green's functions in the presence of a VEV for the Higgs--$A_5$. 
To account for the VEV, we use the well known technique of rotating the $\left<A_5\right>$ into the IR boundary conditions. 
We do this by the bulk gauge transformation $A_\mu \rightarrow e^{ig\int_R^z\,\left<A_5\right>dz'}A_\mu e^{-ig\int_R^z\,\left<A_5\right>dz'}$ to the left of the IR brane and $A_\mu \rightarrow e^{ig_5\int_{R'}^z\,\left<A_5\right>dz'}A_\mu e^{-ig_5\int_{R'}^z\,\left<A_5\right>dz'}$ to the right of the IR brane. 
This eliminates the bulk $\left<A_5\right>$, but changes the IR matching conditions for the $SO(4)\times U(1)_X$ fields to
\begin{eqnarray}
G|_{z=R'^+}\,=\,U(h)\,G|_{z=R'^-}\,U(h)^{-1} \,, ~~~~\plz G|_{z=R'^+}\,=\,U(h)\,\plz G|_{z=R'^-}\,U(h)^{-1}\, .
\end{eqnarray}
where $U(h)~\equiv~e^{ig_5\int_{R}^{R'}\,\left<A_5\right>dz'}\,\equiv\,e^{i g_5 \frac{h}{f}}$. 
Note that the rotation to the right of the IR does not have any impact on the boundary conditions because the $A_5$ is pure gauge in this region.

\subsubsection{Results}

\begin{figure}
\begin{center}
\includegraphics[width=0.6\textwidth]{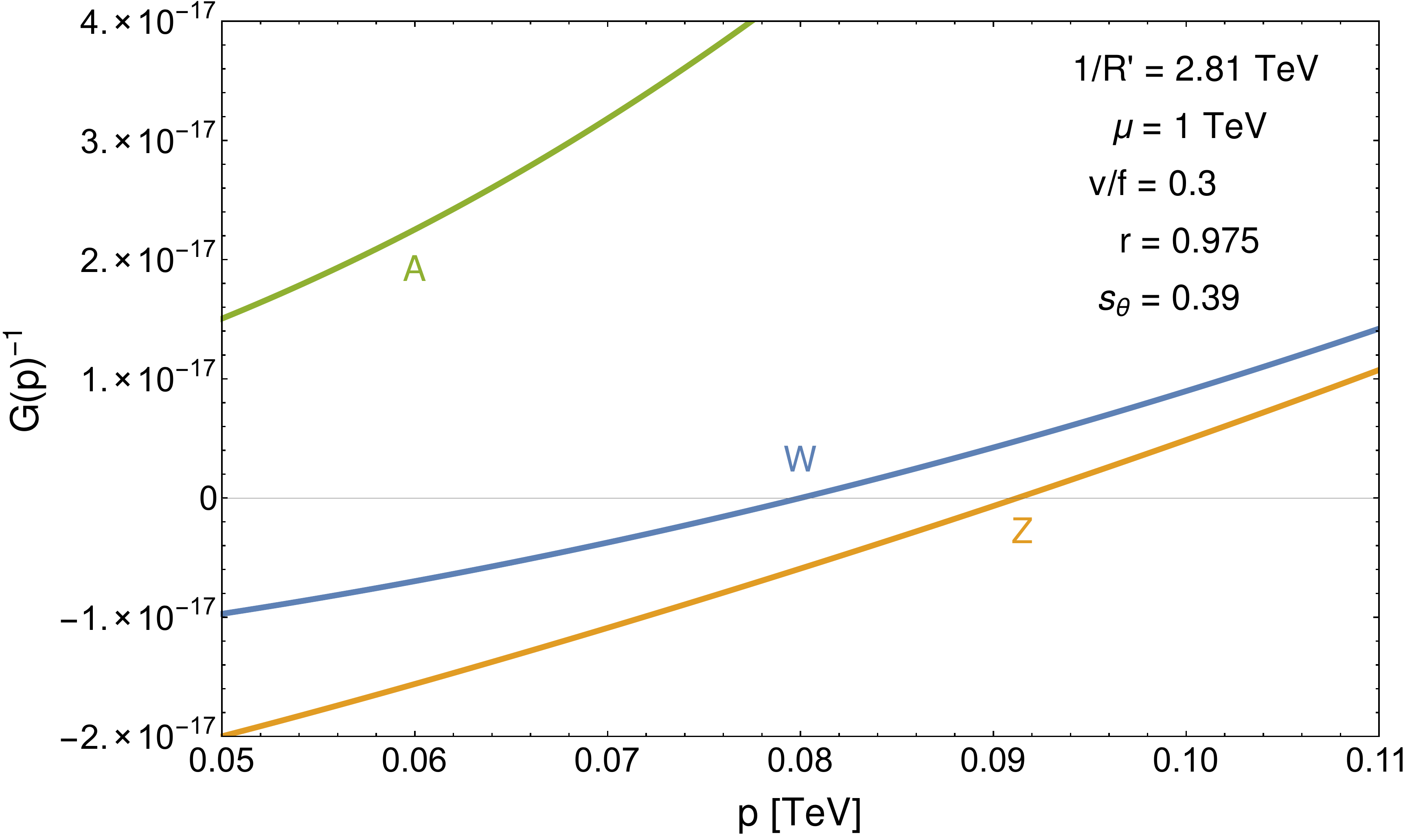}
\end{center}
\caption{Inverse Green's function for the gauge bosons for $v/f=0.3$. The zero modes are at $m_W$ and $m_Z$.}
\label{fig:fullgi}
\end{figure}

In Fig.~\ref{fig:fullg} we present the gauge spectral density, for a Higgs VEV ratio $v/f=0.3$. 
The spectral density is nonzero above the gap $\mu$. 
Note the poles on top of the continuum at $11$~TeV: these are the result of the IR brane Dirichlet B.C. for the generators of $SO(5)/SO(4)$, and are the only BSM poles that appear in our model. 
Since these are well beyond the reach of the LHC, we will not study them further in this paper. 
In addition to the gauge boson continuum, our gauge boson Green's functions reproduce the SM $W$ and $Z$ masses and a massless photon: in Fig.~\ref{fig:fullgi}, we show the inverse of the gauge boson Green's functions, which intersect zero exactly at the SM values.

\subsection{Fermion Spectral Densities}\label{sed:fermspec}

The quadratic fermion Lagrangian is the one from Eq.~\ref{eq:Lagferm}, where each one of the bulk multiplets $Q_L,\,T_R$ and $B_R$ has a different bulk mass $c_Q,\,c_T$ or $c_B$. 
The three multiplets have dilaton Yukawa couplings $y,-y,-y$. 
Note the flip in the sign of the dilaton Yukawa, required to get a zero mode in the $\psi$ component of $T_R,\,B_R$. 
The homogeneous EOM for $Q_L$ are
\begin{equation}\label{eq:bulkeqL} \begin{split}
-\mathcal{D}_f \chi_{Q_L}-\frac{\hat{c}_Q(z) a(z)}{R}\chi_{Q_L}~+~p \,\psi_{Q_L}~&=~0 \,, \\
\mathcal{D}_f \psi_{Q_L}-\frac{\hat{c}_Q(z) a(z)}{R}\psi_{Q_L}~+~p \,\chi_{Q_L}~&=~0 \,,
\end{split} \end{equation}
where $\mathcal{D}_f f~=~\plz f+2\frac{\plz a(z)}{a(z)} f$, $a(z)=\frac{R}{z}e^{-\frac{2}{3}\mu(z-R)}$ and $\hat{c}_Q(z)=(c_Q+\mu(z-R))e^{\frac{2}{3}\mu(z-R)}$.
The general solutions for $Q_L$  are:
\begin{equation}\label{eq:bulksol} \begin{split}
\chi_{Q_L}(z)~&=~a(z)^{-2}\,\left[A \,\hat{M}(c_Q,z)~+~B \,\hat{W}(c_Q,z)\right] \,, \\
\psi_{Q_L}(z)~&=~a(z)^{-2}\,\left[A ~\alpha(c_Q,p)\,\hat{M}(c_Q,z)~+~B ~\beta(p)\,\hat{W}(c_Q,z)\right]\, \,,
\end{split} \end{equation}
with $\alpha(c_Q,p)\equiv\frac{4\left(\frac{1}{2}+c_Q-R\mu y\right)\Delta}{p}$, $\beta(p)\equiv\frac{\mu y-\Delta}{p}$ and $\Delta\equiv\sqrt{\mu^2y^2-p^2}$. 
The functions $\hat{W}(c_L,z),\,\hat{M}(c_L,z)$ are defined as:
\begin{equation} \begin{split}
\hat{M}(c_Q,z)~&=~M\left(\frac{-\mu y \left(c_Q-\mu y R\right)}{\Delta},\,\frac{1}{2}+c_Q-R\mu y, \, 2\Delta z\right) \,, \\
\hat{W}(c_Q,z)~&=~W\left(\frac{-\mu y \left(c_Q-\mu y R\right)}{\Delta},\,\frac{1}{2}+c_Q-R\mu y, \, 2\Delta z\right)\,,
\end{split} \end{equation}
where $M(a,b,z), W(a,b,z)$ are Whittaker functions.
The homogeneous solution for $T_R$ $(B_R)$ is the same as the one for $Q_L$ under $c_{Q}\rightarrow -c_{T(B)}$ and $\chi_{Q_L},\psi_{Q_L} \rightarrow \psi_{T_R(B_R)},\chi_{T_R(B_R)}$.

To find the Green's functions, we have to solve the inhomogeneous version of Eq.~\ref{eq:bulkeqL}, inserting $\delta(z-z')$ on the right side of the first or second of these equation, depending on the Green's function. 
The full details of the calculation are presented in Appendix~\ref{app:fermion}.

The fermionic Green's functions are subject to the boundary conditions given below.

\subsubsection{UV Boundary Conditions}

We assign UV brane Dirichlet B.C. to the following states:
\begin{eqnarray}\label{eq:fUVbc}
Q_L&:&~~\psi_{q_L},\,\chi_{\tilde{q}_L},\,\chi_{y_L} \,, \nonumber\\
T_R&:&~~\psi_{q_R},\,\psi_{\tilde{q}_R},\,\chi_{t_R} \,, \\
B_R&:&~~\psi_{q'_R},\,\psi_{\tilde{q}'_R},\,\psi_{x_R},\,\psi_{y_R},\,\psi_{\tilde{y}_R},\,\chi_{b_R} \,. \nonumber
\end{eqnarray}
These boundary conditions reflect the fact that only $q_L,\,q_R$ and $b_R$ are partially composite and should have the SM fermions as zero modes.

\subsubsection{IR Matching Conditions}

Due to the masses $M_{1,4,b}$ on the IR brane (Eq.~\ref{eq:IRlag}), we have the jump conditions relating $G^{\pm}=G|_{z=R'^+}\pm G|_{z=R'^-}$ for the different multiplets.
\begin{eqnarray}\label{eq:fIRR}
G^-_{t_{R}}~&=&~\kappa M_1\,G^+_{s_{L}},~~~~~~G^-_{y_{L}}~=~\kappa M_1\,G^+_{t_{R}}\,, \nonumber \\
G^-_{a_{R}}~&=&~\kappa M_4\,G^+_{a_{L}},~~~~~~G^-_{a_{L}}~=~\kappa M_4\,G^+_{a_{R}} +\kappa M_b\,G^+_{a'_{R}}\,, \\
G^-_{a'_{R}}~&=&~\kappa M_b\,G^+_{a_{L}} \,, \nonumber
\end{eqnarray}
where $a=q,\tilde{q}$ and $\kappa=\mp 1$ for the $\chi/\psi$ Green's function. 
We only select Green's functions that are regular in the deep IR, as described in Sec.~\ref{sec:LD}. 

\subsubsection{Accounting for the VEV of the Higgs--$A_5$}

Similarly to the gauge case, we can account for the VEV of the Higgs--$A_5$ by acting on the fermion with the Wilson line $U(h)$. 
The IR matching conditions remain the same as long as we modify the definition of $G^{\pm}$ to be
\begin{eqnarray}
G^{\pm}~=~G^h|_{z=R'^+}\,\pm\,G^h|_{z=R'^-}\, ,
\end{eqnarray}
where $G^h~=~U(h)G$ for $Q_L,\,T_R$ in the $\mathbf{5}$ of $SO(5)$ and $G^h~=~U(h)\,G\,U(h)^{-1}$ for $B_R$ in the $\mathbf{10}$ of $SO(5)$.

\subsubsection{Results}

\begin{figure}
\begin{center}
\includegraphics[width=0.6\textwidth]{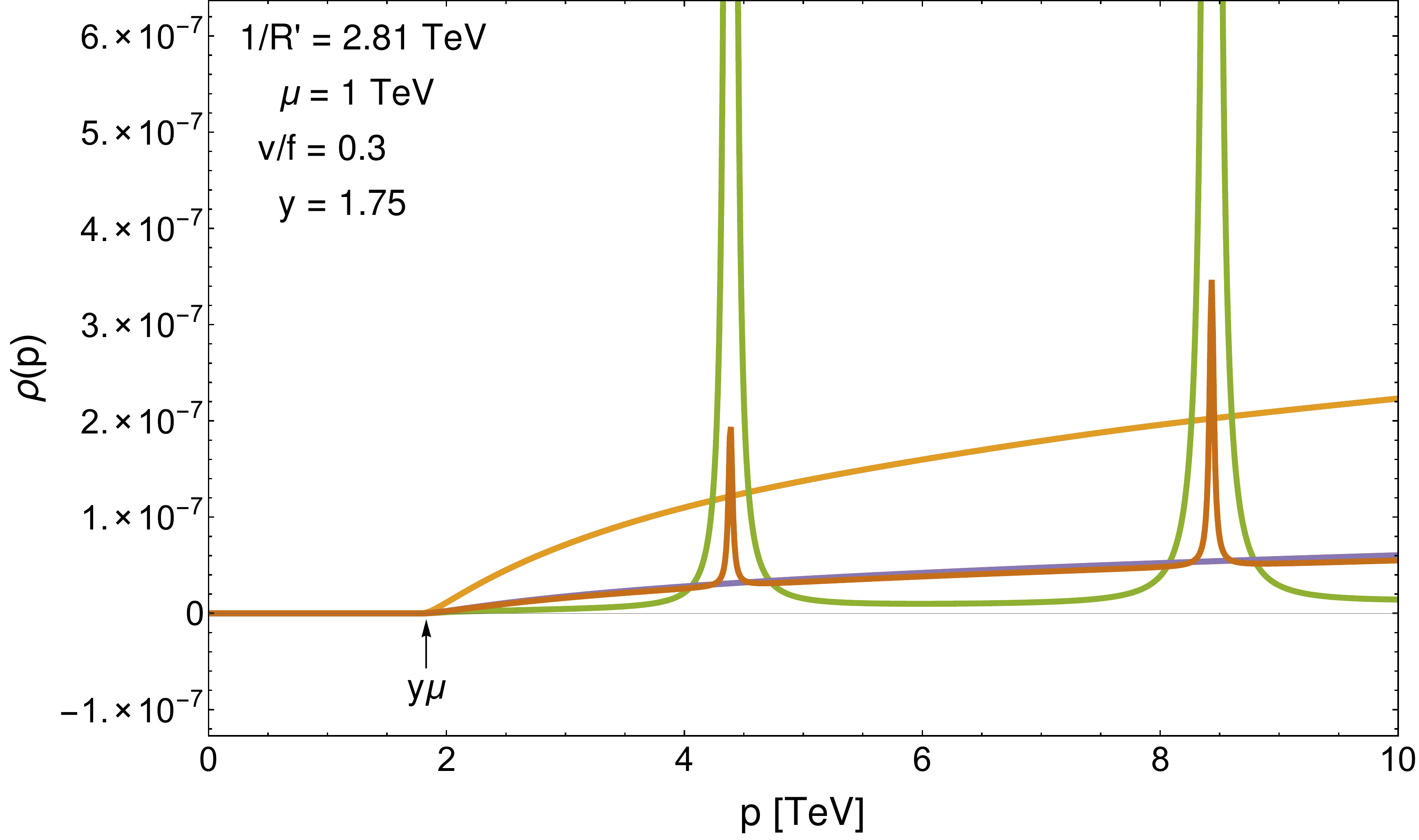}
\end{center}
\caption{Spectral densities for additional exotic top partners.}
\label{fig:fullf2}
\end{figure}

\begin{figure}
\begin{center}
\includegraphics[width=0.6\textwidth]{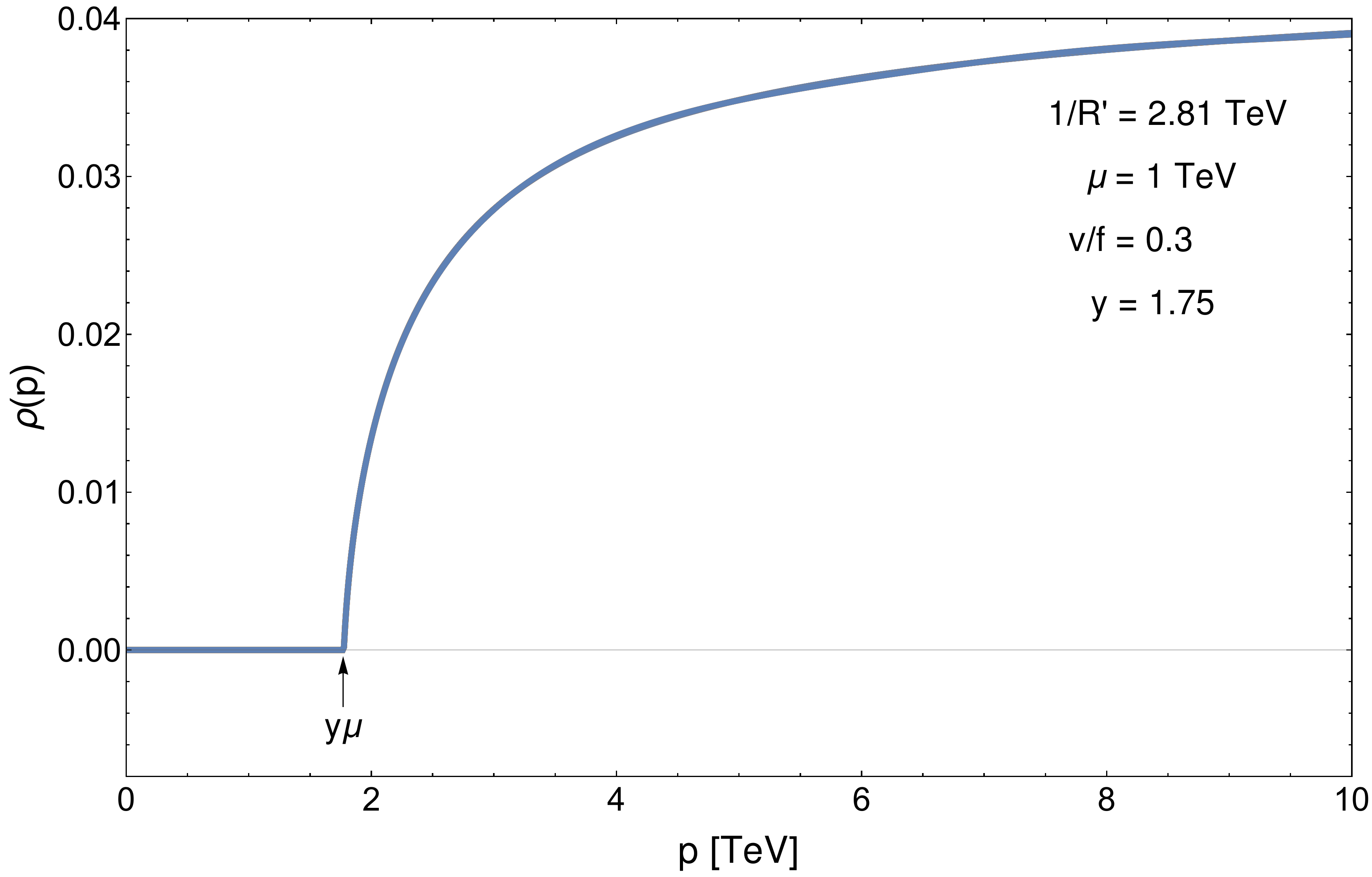}
\end{center}
\caption{Spectral densities for the remaining top partner quantum numbers. 
The figure contains ten overlapping spectral densities corresponding to components that are continuous across the IR brane.}
\label{fig:fullf3}
\end{figure}

Here we show the final results for the fermionic spectral densities for the parameter choices in Eq.~\ref{eq:benchmark}. 
Since the bulk matter content of the full model is given by the representations $\mathbf{5}+\mathbf{5}+\mathbf{10}$, we will end up with a $20\times 20$ matrix for the Green's functions. 
Each entry of the matrix is a two-point function between a pair of fermions, and diagonalizing this yields the 20 fermionic spectral densities of our model.
In Fig.~\ref{fig:fullf} we can see the spectral densities for $t,b$ and an exotic $b'$. 
All other spectral densities corresponding to states with other quantum numbers in the decomposition Eq.~\ref{eq:decomp} are shown in Figs.~\ref{fig:fullf},~\ref{fig:fullf2}, and~\ref{fig:fullf3}. Note that the difference in normalization between the different spectral densities stems from the difference in bulk mass between the three bulk multiplets, which leads to factors of ${\left(\frac{R}{R'}\right)^{\Delta c}}$ between the spectral densities depicted in the figures.

\begin{figure}
\begin{center}
\includegraphics[width=0.6\textwidth]{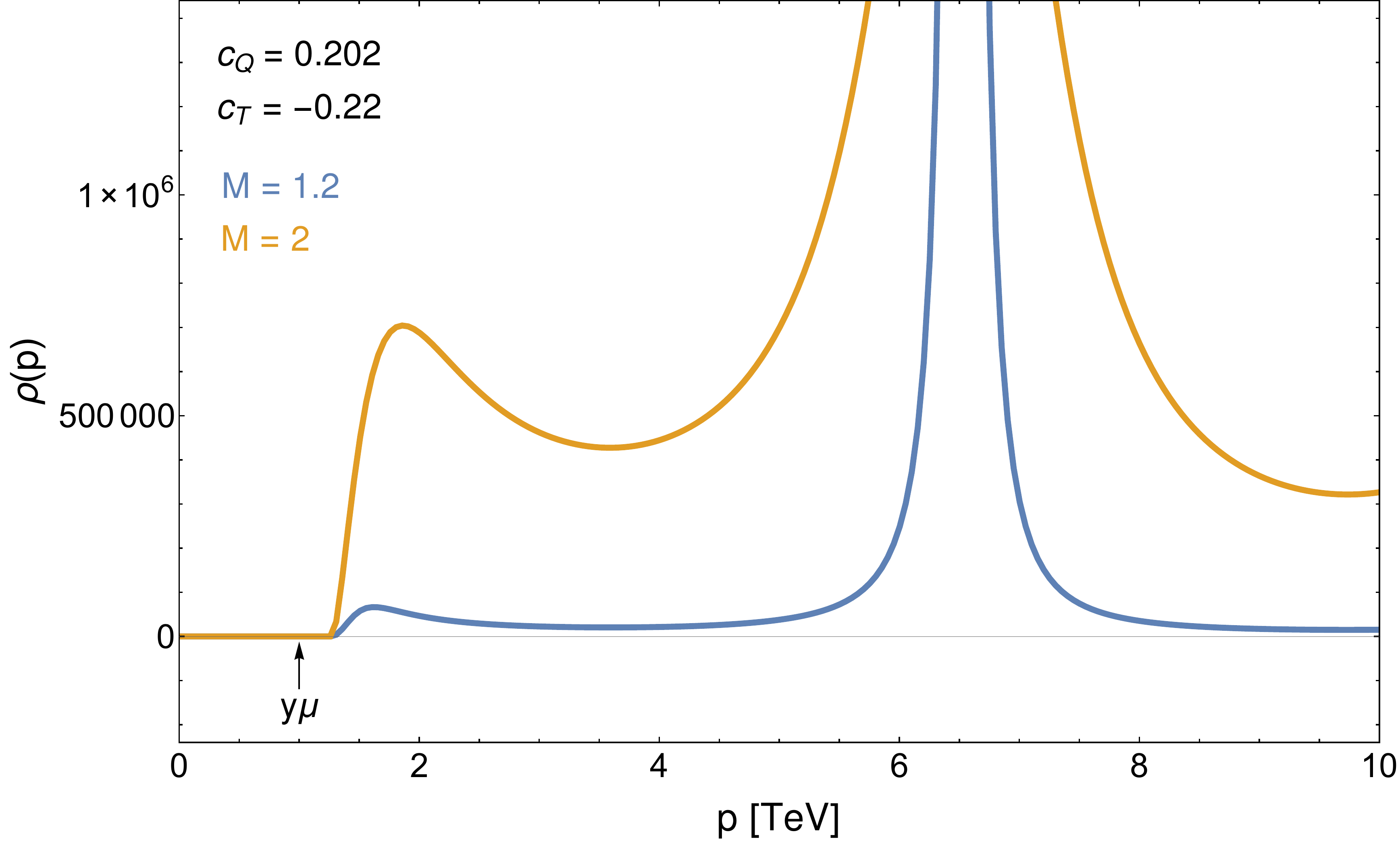}
\end{center}
\caption{The effect of the IR mass on the width of a fermionic peak in a toy model with a single bulk fermion. By varying the IR mass, the peak could be made as wide as $2$~TeV.}
\label{fig:width}
\end{figure}

As stated before, some of the spectral functions feature broad peaks that can be probed at future colliders. The width of these peaks is a model dependent parameter which depends on the magnitude of the IR mass $M_1$. By choosing a slightly different point in parameter space with $M_1=2$, we can make these peaks as wide as $2$~TeV. This effect is depicted in Fig~\ref{fig:width} in a toy model with a single bulk fermion.

\begin{figure}
\begin{center}
\includegraphics[width=0.6\textwidth]{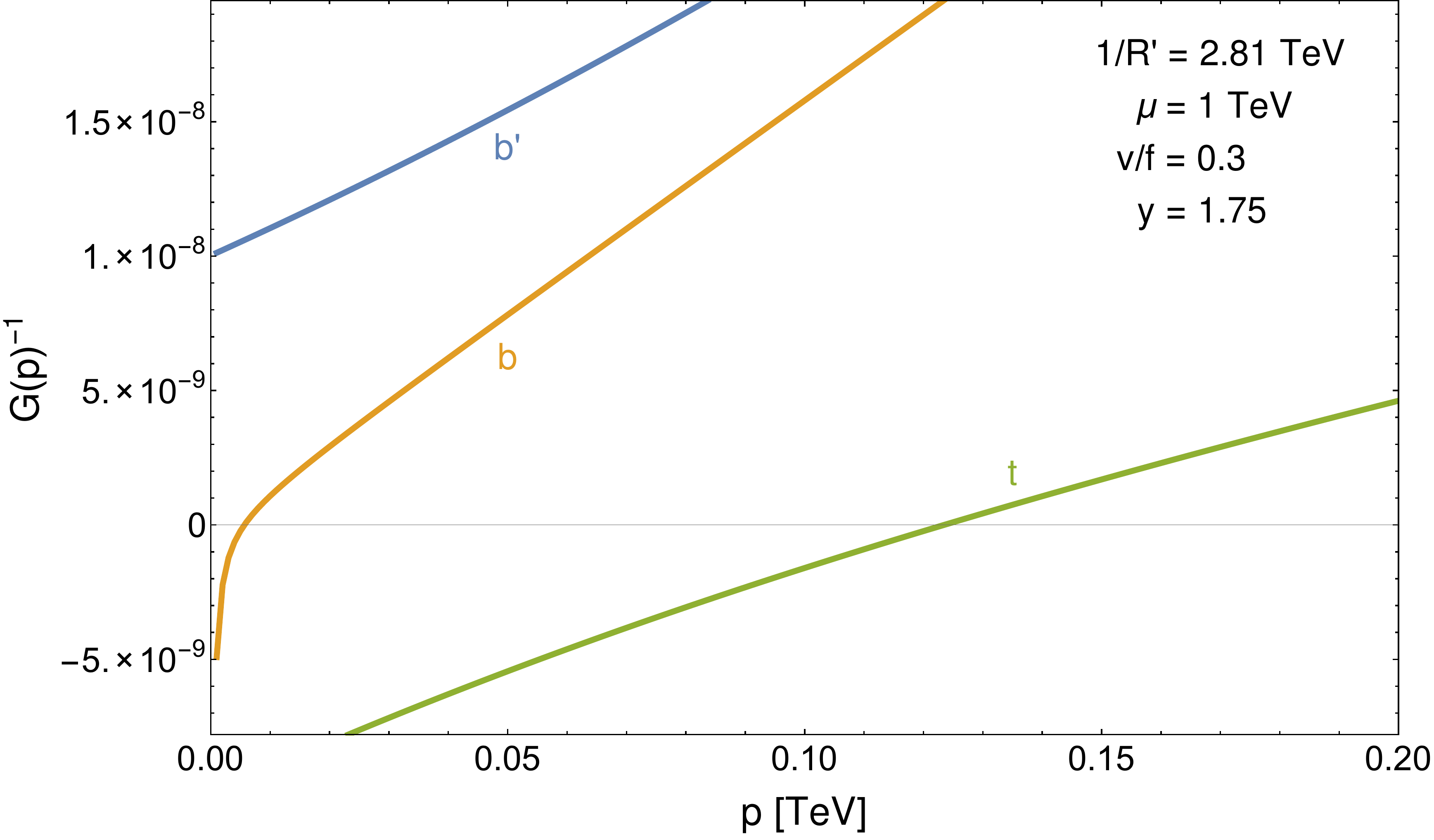}
\end{center}
\caption{Top, bottom and $b'$ inverse Green's functions. The zero modes of $t$ and $b$ are lifted in the presence of the Higgs VEV.}
\label{fig:fullfi}
\end{figure}

Without a Higgs VEV, four of the 20 fermions, $t_{L,R}$ and $b_{L,R}$, would have zero modes.
The Higgs VEV lifts these to $m_t$ and $m_b$, as shown in the inverse Green's functions for $t, b$ and $b'$ in Fig.~\ref{fig:fullfi}.

\newpage
\section{The Higgs Potential}~\label{sec:Higgspot}

Given all of the gauge and fermion Green's functions that we have calculated, it is straightforward to compute the Coleman-Weinberg potential for the Higgs using the formula \cite{Csaki:2008zd}:
\begin{eqnarray}
V(h)~=~\frac{3}{16\pi^2}~\int\,dp\,p^3\,\left[-4\sum_{j=1}^{20}\,\log G_{f_j}(ip)~+~\sum_{k=1}^{4}\,\log G_{g_k}(ip)\right]\, ,
\end{eqnarray}
where $G_{f_j}(p)$ and $G_{f_k}(p)$ are the eigenvalues of the fermion and gauge Green's function matrices, respectively. 
Note that these Green's functions are Higgs-dependent, hence their contribution to the Coleman-Weinberg potential. 
In Fig.~\ref{fig:Higgspot} we plot the Coleman-Weinberg potential as a function of $\sin \left<h\right>/f \equiv v/f$. 
There is a minimum at $v/f=0.3$, and so the right Higgs VEV is obtained for $f=820$~GeV, which is consistent with electroweak precision bounds. 
By differentiating this twice, one can show that indeed $m_h=125$~GeV. 

By varying the parameters of our model, we can estimate the tuning for the BP. 
We use the Barbieri-Giudice measure to quantify the tuning:
\begin{eqnarray}
\text{tuning}~=~\left[\text{max}_i\,\frac{d\log\,v}{d\log\,p_i}\right]^{-1}\, ,
\end{eqnarray}
where $p_i\in\left\{R,R',\mu,r,\theta,y,c_Q,c_T,c_B,M_1,m_4,M_b\right\}$ are the fundamental parameters of the model. 
We obtain a tuning of $1\%$ for the BP, with the strongest dependence being on $c_Q$ and $c_T$ as expected.

\section{Comments on Phenomenology} \label{sec:pheno}

The detailed study of continuum partner phenomenology will appear in a separate work \cite{CLLT}. 
Here, we will merely point out some the main points regarding the phenomenology of continuum partners.

\begin{enumerate}

\item \textit{No s-channel resonances at the LHC:} 
The unique feature of our continuum CH model is the lack of particle resonances within the reach of the LHC. 
This leads to vastly different phenomenology, in which the traditional searches for KK gauge bosons no longer apply, as well as all of the resonance-based top partner searches. 
To demonstrate this point, we present the partonic cross-section $\sigma\left(q\bar{q}\rightarrow G^* \rightarrow t_R\bar{t}_R\right)$, in three simplified models: only SM gluon, KK gluons, and continuum gluons. 
This cross section is given by
\begin{eqnarray}
\sigma(\hat{s})~=~\sigma(\hat{s})_{\text{SM}}\,\times\,{\hat{s}}^2\,{\left|G(R,R';\hat{s})\right|}^2\, ,
\end{eqnarray}
where $G(R,R';\hat{s})$ is the UV to IR Green's function, calculated in a similar manner to \secref{specden}. 
The results are depicted in Fig.~\ref{fig:schannel}.

\item \textit{Bounds from the running of $\alpha_{S}$:}
The running of $\alpha_s$ in the presence of a colored gauge boson continuum provides an interesting bound on $\mu$, the starting point of the gapped continuum.
The running of the 4D gauge coupling is given by~\cite{ArkaniHamed:2000ds,Contino:2002kc,Csaki:2008zd}
\eql{RGbraneloc}{
\frac1{g^2(Q)} = \frac1{g_5^2} \int_{R}^{1/Q} dz \, a(z) + \frac1{g\ld{UV}^2} - \frac{b\ld{UV}}{8\pi^2} \log\left(\frac1{RQ}\right) \,,
}
where $g_5^2 = g_*^2 R$, $g\ld{UV}$ is the UV brane coupling, and $b\ld{UV}$ is the one-loop beta function including effects for the zero modes on the UV brane.
In our case, we localize all fields except for $t_R$ on the UV brane, so $b\ld{UV} = 22/3$.

There are determinations of $\alpha_s$ up to $Q \sim 1.42$~TeV from measurements of jets by CMS using $\sqrt{s} = 7$~TeV LHC data \cite{Chatrchyan:2013txa,CMS:2014mna,Khachatryan:2014waa,Patrignani:2016xqp}.
For a given value of $\mu_g$, we choose the SM value for $\alpha_s(\mu_g)$.
Evaluating \eqnref{RGbraneloc} at two different scales $Q = \mu_g, 1.42$~TeV and taking the difference, we can determine the effect of the continuum on the running of $\alpha_s$.
Results with different UV-brane localized values of $\alpha\ld{UV}$ are shown in Fig.~\ref{fig:alphasRG}: gluon continuum scales above $\mu = 600$--$700$~GeV are generically safe from the $1\sigma$-high value of $\alpha_s(1.42$~TeV).
The value of $\alpha\ld{UV} = 0.025$ corresponds to the limit $g_*^2 < (4\pi)^2$ from requiring perturbativity in the bulk.

\begin{figure}[htb]
\centering
\includegraphics[width=0.6\textwidth]{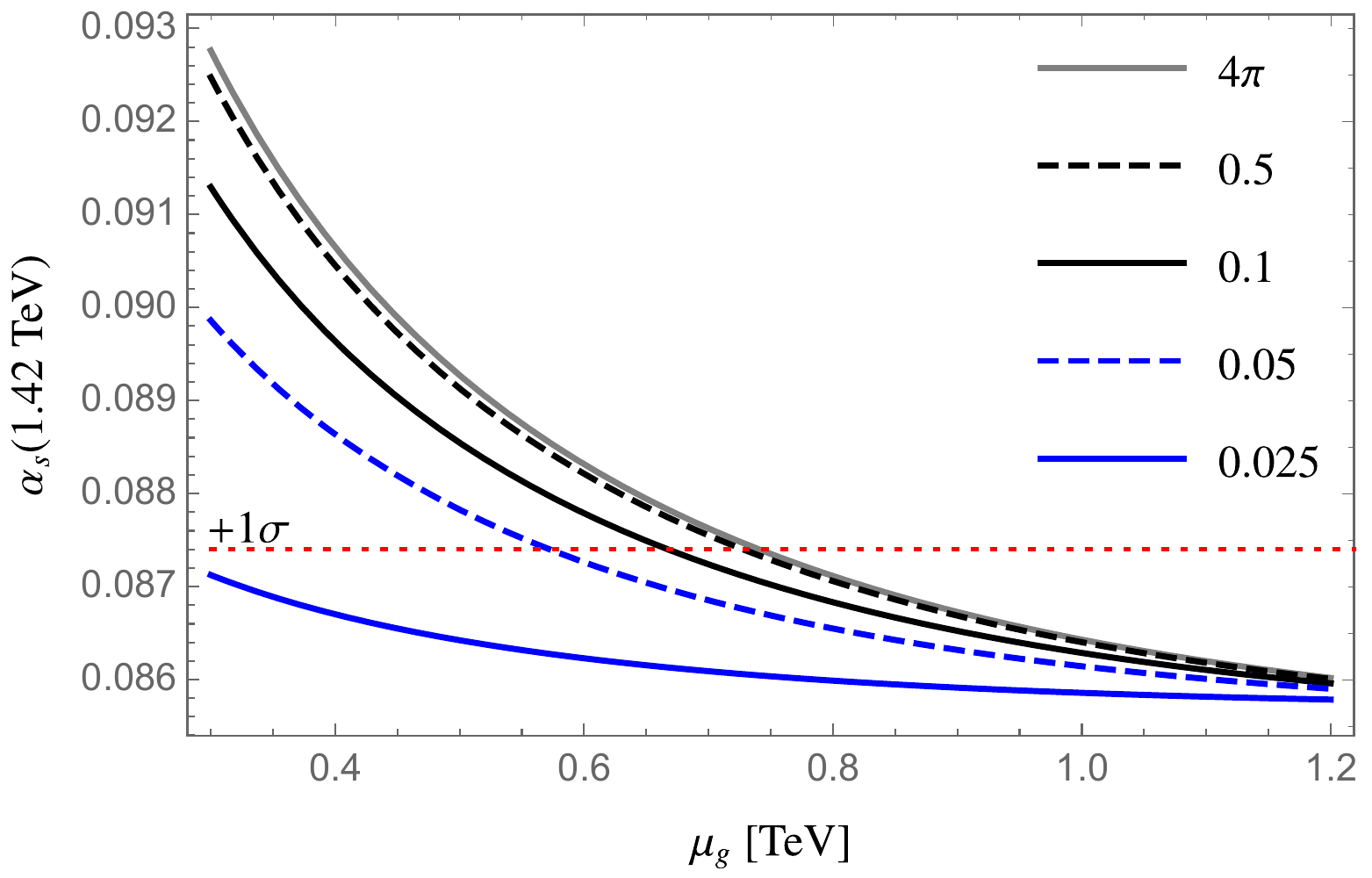}
\caption{Bound from running $\alpha_s$ from the gluon continuum scale $\mu_g$ to $Q = 1.42$ TeV.}
\label{fig:alphasRG}
\end{figure}

\item \textit{Pair production of continuum top partners:}
We expect the continuum top pair-production cross section to be parametrically smaller than the one for particle top partners. 
This is due to the smearing of the spectral density to higher energies, where PDF suppression dominates. 
This is also the case in the result of \cite{Cacciapaglia:2008ns} for colored fermionic unparticles. 

The full calculation of the pair-production cross section for continuum top partners is far from trivial, as it is unclear how to calculate the phase space factor for a pair of final-state continuum fermions. 
Inspired by the work of \cite{Cacciapaglia:2007jq,Cacciapaglia:2008ns}, we can use the optical theorem to relate the pair-production cross section to the imaginary part of the diagram of the vacuum polarization with a continuum fermion loop, which we can calculate using dispersion relations. 
We leave the full calculation of the top partner pair production for future work \cite{CLLT}. 

Since we are not calculating the full cross section in this paper, we chose a conservative point in parameter space with a continuum fermion gap of $1.75$~TeV, with $1\%$ tuning.

\end{enumerate}

\section{Conclusions}\label{sec:concl}

We presented a novel type of composite Higgs model, where all top and gauge partners form continua rather than being ordinary particles. 
Such top and gauge partners will evade all $s$-channel resonance searches and are expected to lead to unique experimental signatures.
We showed how to obtain a realistic model of this sort from a warped extra dimension, where space continues beyond the IR brane. 
A linear dilaton dominates the deep IR region, corresponding to critical IR dynamics that produces a gapped continuum. 
We have shown how to calculate the full set of spectral densities for the fermion and gauge partners. 
Furthermore, we established a phenomenologically viable benchmark point, tuned at the percent level, with a realistic radiatively generated Higgs potential. 
The phenomenology of continuum partners, as well as the existing collider bounds on this model, will appear in an upcoming publication.

\section*{Acknowledgments}

The authors are grateful for conversations with Mihailo Backovic, Steven Durr, Michael Geller, and Jesse Thaler.
The authors are also grateful for the Mainz Institute for Theoretical Physics (MITP) for its hospitality and its partial support while this work was in progress.
C.C., G.L., and S.J.L. thank the Aspen Center for Physics, which is supported by National Science Foundation grant PHY-1607611, for its hospitality and its support.
C.C., G.L., S.L. and O.T. are supported by the U.S. National Science Foundation through grant PHY-1719877.
G.L. and S.J.L. acknowledge support by the Samsung Science and Technology Foundation under Project Number SSTF-BA1601-07, and a grant from Korea University. C.C. is supported in part by the US-Israel BSF grant 2016153.
This work was partially supported by a grant from the Simons Foundation.


\begin{appendix}
\section*{Appendix}

\section{Gauge Boson Green's Functions\label{app:gauge}}

In this appendix we present the details necessary for evaluating the gauge boson spectral densities. We demonstrate our calculation in a simple toy model with a bulk $U(1)\times U(1)$ broken to $U(1)$ on the IR brane, where the role of the Higgs is played by the $A_5$ of the broken $U(1)$. In the bulk we have two gauge bosons, $W_\mu$ and $W'_\mu$. $W_\mu$ corresponds to an unbroken direction (before the Higgs VEV)  with the boundary conditions $\left(+,+\right)$ on the UV and IR branes, while $W'_\mu$ corresponds to a broken direction $\left(-,-\right)$. The Higgs VEV will mix these two fields. 
There are now four different Green's functions to solve for: $\left<W(z)W(z')\right>,\,\left<W'(z)W(z')\right>,\,\left<W(z)W'(z')\right>$ and $\left<W'(z)W'(z')\right>$, which we denote collectively by $G(z,z';p^2)_{ij}, i,j\in\{W,W'\}$.

To get each spectral density we need to first specify whether the source (the delta function) is in the $W$ or the $W'$ equation. This depends if we're looking for $W(z')$ or $W'(z')$ in the correlation function. In a shorthand notation, we will write
\begin{eqnarray}\label{eq:gg2}
\mathcal{D}_g\,G(z,z';p^2)_{ij}~=~\d_{ij}\, \delta(z-z') \,.
\end{eqnarray}

\begin{figure}
\begin{center}
\includegraphics[width=0.6\textwidth]{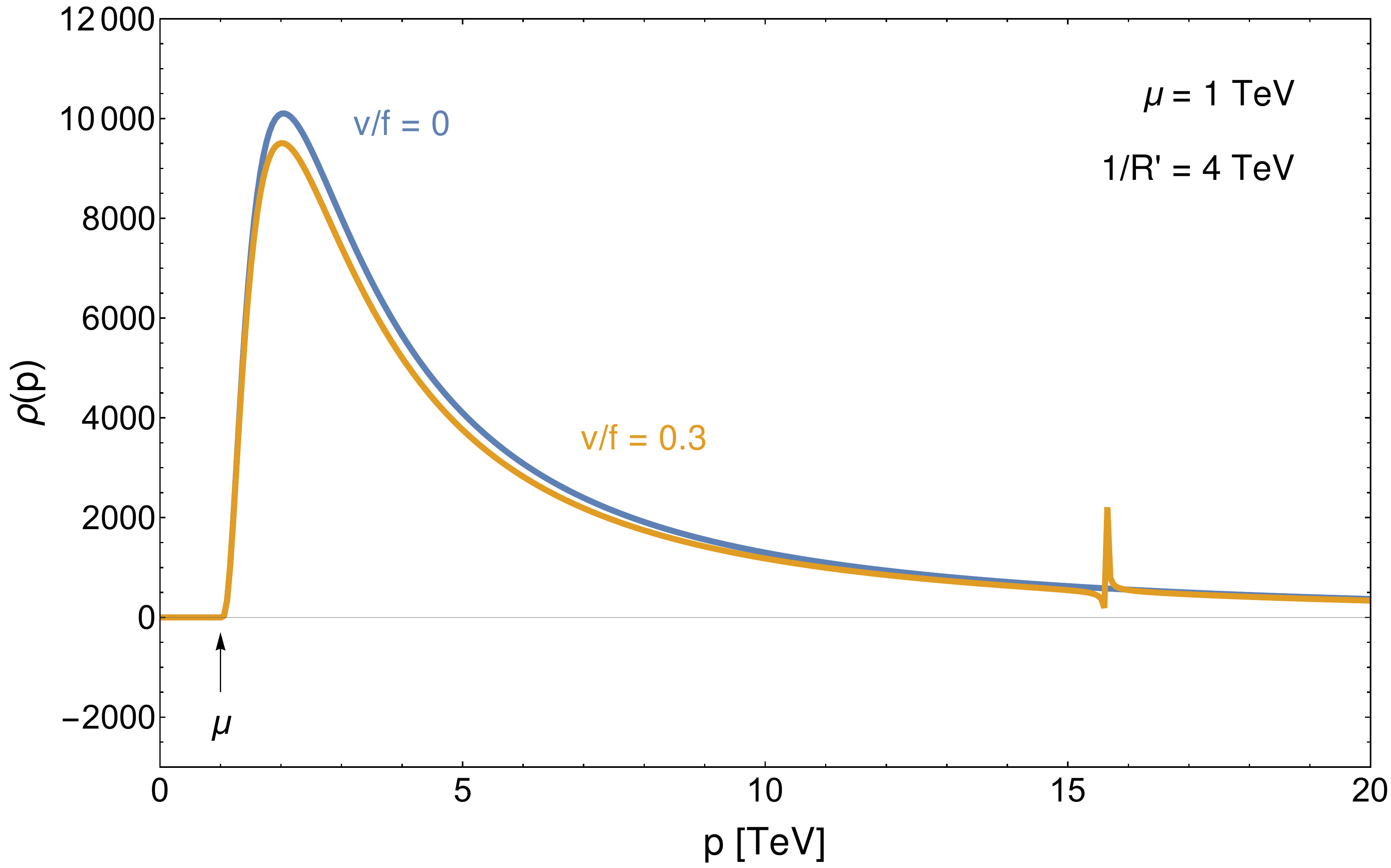}
\end{center}
\caption{Spectral density for $W/W'$ mix with $\mu=1\,\text{TeV},\,1/R'=4$~TeV.}
\label{fig:2g}
\end{figure}

\begin{figure}
\begin{center}
\includegraphics[width=0.6\textwidth]{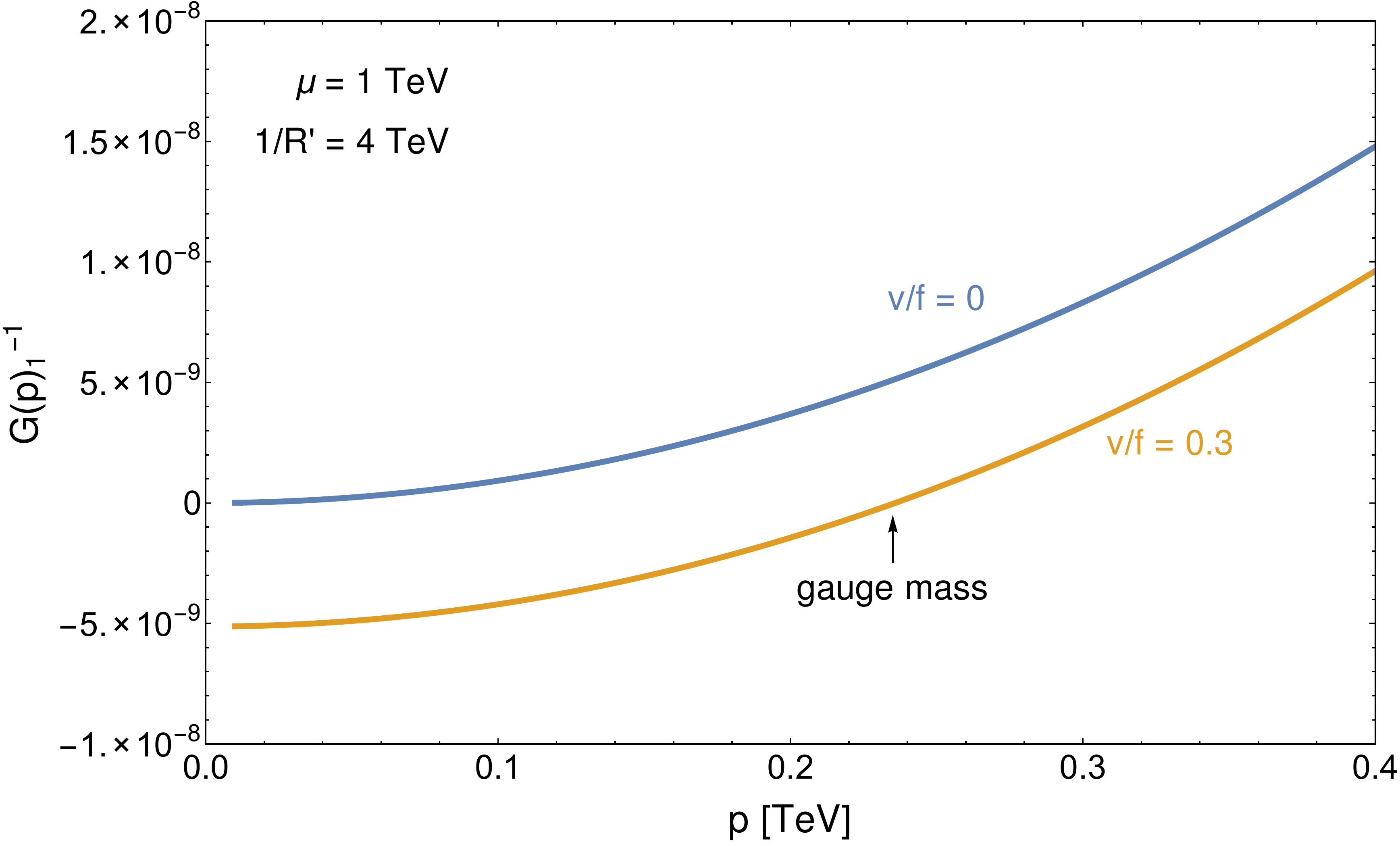}
\end{center}
\caption{Inverse Green's function for $W/W'$ mix with $\mu=1\,\text{TeV},\,1/R'=4$~TeV.}
\label{fig:wmass}
\end{figure}

The gauge EOM is Eq.~\ref{eq:gaugeeomd}, with a general solution of the form Eq.~\ref{eq:bulksolg}.
To solve for the Green's function, we divide our space into the domains $R\leq z\leq z',\,z'\leq z\leq R'$ and $R'\leq z$, where the coefficients are denoted $\ola{A},\ola{B}$, $\ora{A},\ora{B}$ and $A^{\infty},B^{\infty}$, respectively. Our goal is then to find these coefficients, subject to the following jump/boundary conditions:
\begin{itemize}
\item UV BC:
\begin{eqnarray}
\plz G_{Wj}|_{z=R}~=~G_{W'j}|_{z=R}~=~0\, .
\end{eqnarray}
\item Jump conditions at $z=z'$:
\begin{equation} \begin{split}
\Delta G_{Wj}|_{z=z'}~&=~\Delta G_{W'j}|_{z=z'}~=~0 \,, \\
\Delta \plz G_{ij}|_{z=z'}~&=~a(z)^{-1} \d_{ij} \,.
\end{split} \end{equation}
\item Jump conditions at $z=R'$:
\begin{equation} \begin{split}
\Delta G_{Wj}|_{z=R'}~&=~\Delta \plz G_{Wj}|_{z=R'}~=~0\,, \\
G_{W'j}|_{z=R'^-}~&=~G_{W'j}|_{z=R'_+}~=~0 \,. 
\end{split} \end{equation}
\item Regularity of $G(z,z,p^2)$ at $z\rightarrow \infty$:
\begin{eqnarray}
A^\infty_{Wj}=A^\infty_{W'j}=0\, .
\end{eqnarray}
\end{itemize}
From these linear conditions we arrive at the Green's function matrix $G(z,z';p^2)_{ij}$. 
We can of course diagonalize this matrix and define two spectral densities:
\begin{eqnarray}
\rho(s)_{1,2}~\equiv~\lim_{z,z'\rightarrow R}\,\frac{1}{\pi}\text{Im}~G(z,z';s)_{1,2}\, .
\end{eqnarray}
corresponding to the two eigenvalues of $G_{ij}$. 
Note, however, that in the present case $G_{W'W}=G_{W'W'}=0$ by virtue of the UV Dirichlet BC, and also that $G_{WW'}=0$ because the $W$ and $W'$ IR BC are completely decoupled. 
The situation is different once we consider a bulk $\left<A_5\right>$ VEV rotating the two multiplets. 
In this case the IR BC is modified to:
\begin{equation} \begin{split}
\Delta \hat{G}_{Wj}|_{z=R'}~&=~\Delta \plz \hat{G}_{Wj}|_{z=R'}~=~0\,, \\
\hat{G}_{W'j}|_{z=R'^-}~&=~\hat{G}_{W'j}|_{z=R'_+}~=~0\,,
\end{split} \end{equation}
with
\begin{equation} \begin{split}
\hat{G}_{Wj}~&=~~~~~c_h\,G_{Wj}~+~s_h\,G_{W'j} \,, \\
\hat{G}_{W'j}~&=~-s_h\,G_{Wj}~+~c_h\,G_{W'j} \, ,
\end{split} \end{equation}
and $c_h=\cos \left(\left<h\right>/f\right),\,s_h=\sin \left(\left<h\right>/f\right)\equiv v/f$. 
The resulting spectral density $\rho_1(p)$ is depicted in Fig.~\ref{fig:2g} ($\rho_2(p)$ is still zero by virtue of the Dirichlet BC for $W'$). 
Note that for $v/f\neq 0$, $\rho_1(p)$ acquired a pole on top of the branch cut. 
This is as expected: the pole originates from the $W'$ component of the eigenstate $1$, and $W'$ has a Dirichlet BC at $R'$. 
We also note that the zero mode of eigenstate $1$ is no longer massless since it acquires a mass from the Higgs. 
This is seen by determining where $1/G_1(p)$ vanishes, as shown in Fig.~\ref{fig:wmass}.

We are now ready to present the spectral densities for the gauge bosons of the realistic model. 
As previously explained, the bulk gauge symmetry $SO(5)\times U(1)_X$ is broken on the IR brane to $SO(4)\times U(1)_X$. 
We choose the following generators as a basis of $SO(5)$:
\begin{equation} \begin{split}
\left(T^a_{L,R}\right)_{ij}~&=~-\frac{i}{2}\left[\,\epsilon^{abc}\delta^b_i\delta^c_j~\pm~\left(\delta^a_i\delta^4_j-\delta^4_i\delta^a_j\right)\,\right]\,,\\
\left(T^{\hat{a}}_{C}\right)_{ij}~&=~-\frac{i}{\sqrt{2}}\left[\,\left(\delta^{\hat{a}}_i\delta^5_j-\delta^5_i\delta^{\hat{a}}_j\right)\,\right]\, ,
\end{split} \end{equation}
with $a\in\{1,2,3\}$ and $\hat{a}\in\{1,2,3,4\}$. 
The gauge bosons corresponding to the generators $T^a_{L,R}$ are continuous across the IR brane, while the ones corresponding to $T^{\hat{a}}_{C}$ get a Dirichlet BC. 
The Higgs emerges as the $4D$ component of $A^{C\hat{a}}_5$. 
To solve for the bulk profiles in the presence of a Higgs VEV, we use the trick of solving for $v=0$ but with the IR BC rotated by the Higgs matrix: 
\begin{eqnarray}
U_h~=~e^{i\,g_5\,\int_R^{R'}\,A_5(z)T_C^{\hat{a}}\,h^{\hat{a}}(x)\,dz}\, .
\end{eqnarray}
The profile $A_5(z)$ is given by solving the bulk EOM:
\begin{eqnarray}
\partial_z\left[a(z) A_5(z)\right]~=~0\, .
\end{eqnarray}
and so $A_5(z)~=~N_5\,a^{-1}(z)$, with $N_5={\left[\int_R^{R'}\,\,a^{-1}(z)dz\right]}^{-1/2}$. We only integrate $A_5$ up to the IR brane because it is pure gauge beyond it. 
In other words, the only gauge invariant Wilson line for the $A_5$ is between the UV and IR branes.

Due to the SM gauge freedom, we are allowed to choose the Higgs VEV $\left<h\right>$ to be in the $\hat{a}=4$ direction. We then have:
\begin{eqnarray}
U_h~=~e^{i\,T_C^{\hat{4}}\,\left<h\right>/f}\, ,
\end{eqnarray}
where $f=g^{-1}_5{\left[\int_R^{R'}\,\,a^{-1}(z)dz\right]}^{-1/2}$. Substituting the expression for $T_C^{\hat{4}}$, we get
\begin{eqnarray}
U_h~=~\colmatfi{0&0&0&0&0\\ 0&0&0&0&0\\0&0&0&0&0\\ 0&0&0&\cos \left(\left<h\right>/f\right) & \sin \left(\left<h\right>/f\right)\\ 0&0&0&
-\sin \left(\left<h\right>/f\right) & \cos \left(\left<h\right>/f\right)}\, .
\end{eqnarray}

The components of the bulk $SO(5)$ gauge field are:
\begin{eqnarray}\label{eq:gdeco}
W_\mu~=~W^{\pm,3;L}_\mu\,T^{\pm,3}_L~+~W^{\pm,3;R}_\mu\,T^{\pm,3}_R~+~W^{\pm,\pm}_\mu\,T^{\pm\pm}\, ,
\end{eqnarray}
where
\begin{equation} \begin{split}
T^{\pm}_{L,R}~&=~\frac{1}{\sqrt{2}}\left(T^1_{L,R}\pm i T^2_{L,R}\right)\,, \\
T^{\pm\mp}~&=~\frac{1}{\sqrt{2}}\left(T^3_C\pm i T^4_C\right)\,, \\
T^{\pm\pm}~&=~\frac{1}{\sqrt{2}}\left(T^1_C\pm i T^2_C\right)\,.
\end{split} \end{equation}
There is an additional bulk $U(1)_X$ field that we denote by $X_\mu$. 
The extra $U(1)_X$ is required to reproduce the correct hypercharge assignments for the SM fields \cite{Agashe:2004rs}.

The UV boundary conditions are then
\begin{equation} \begin{split}
&W^{\pm;R}_\mu\,|_{z=R}~=~W^{\pm,\pm}_\mu\,|_{z=R}~=~0\,, \\
&\plz\,W^{\pm,3;L}_\mu~+~C_k\,W^{\pm,3;L}_\mu\,|_{z=R}~=~0\,, \\
&c_\theta\,W^{3;R}_\mu~-~s_\theta\,X_\mu\,|_{z=R}~=~0\,, \\
&\plz\left[s_\theta\,W^{3;R}_\mu~+~c_\theta\,X_\mu\right]\,|_{z=R}~=~0 \,.
\end{split} \end{equation}
Note that the above boundary conditions mix the gauge fields $W^{3;R}_\mu$ and $X_\mu$ with an angle $\theta$ as a result of a UV boundary kinetic term. 
A similar boundary kinetic term $C_k\equiv R p^2 r^2 \log R'/R$ is responsible for the mixed boundary conditions for $W^{\pm,3;L}_\mu$. 
This term is necessary to get the right $SU(2)_L$ gauge coupling for every value of the bulk coupling $g_5$. 
To account for the VEV of the Higgs--$A_5$, we rotate its effect into the IR jump conditions, which are now given are given in terms of the Higgs-rotated gauge fields:
\begin{eqnarray}
\hat{W}_\mu~=~U_h\,W_\mu\,U^{-1}_h\, .
\end{eqnarray}
This multiplet is decomposed similarly to Eq.~\ref{eq:gdeco}:
\begin{eqnarray}
\hat{W}_\mu~=~\hat{W}^{\pm,3;L}_\mu\,T^{\pm,3}_L~+~\hat{W}^{\pm,3;R}_\mu\,T^{\pm,3}_R~+~\hat{W}^{\pm,\pm}_\mu\,T^{\pm\pm}\, .
\end{eqnarray}
The IR BC are then
\begin{equation} \begin{split}
&\Delta \hat{W}^{\pm,3;L}_\mu\,|_{z=R'}~=~\Delta \hat{W}^{\pm,3;R}_\mu\,|_{z=R'}~=~0 \,,\\
&\hat{W}^{\pm,\pm}_\mu\,|_{z=R'}~=~0 \,.
\end{split} \end{equation}
The resulting gauge boson spectral densities and inverse Green's functions are presented in the main text in Figs.~\ref{fig:fullg} and~\ref{fig:fullgi}.

\section{Fermion Green's Functions\label{app:fermion}}

Next we explain how an $A_5$ (Higgs) VEV can give mass to two fermion zero modes. 
This is a standard result in gauge-Higgs unification, and we demonstrate it here explicitly in the context of continuum CH.
As a starting point, we take two bulk fermions $Q^{1,2}_L$ with equal bulk masses and dilaton Yukawas $c_L,y$ and opposite UV boundary conditions, and two bulk fermions $Q^{1,2}_R$ with equal bulk masses and dilaton Yukawas $c_R,-y$ and opposite UV boundary conditions.
In the context of warped CH models, $Q^{1,2}_L$ will represent two different quantum numbers within the same $Q_L$ bulk multiplet that is in a large representation of the bulk gauge group, and likewise for $Q^{1,2}_R$. 
However, in this toy example we will not be concerned with the group theory aspects of the model. 
The UV boundary conditions are:
\begin{eqnarray}
\psi^1_L|_{z=R}~=~\chi^2_L|_{z=R}~=~\psi^1_R|_{z=R}~=~\chi^2_R|_{z=R}~=~0\, ,
\end{eqnarray}
where $Q^{1,2}_L=\left(\chi^{1,2}_L,\psi^{1,2}_L\right)$ and $Q^{1,2}_R=\left(\chi^{1,2}_R,\psi^{1,2}_R\right)$.
These boundary conditions, together with the assignment of $\pm y$ dilaton Yukawas and the demand for a regular solution in the deep IR, implies zero modes for $\chi^1_L$ and $\psi^2_R$. 
As IR mass terms only couple $(Q^1_L,Q^1_R)$, $(Q^2_L,Q^2_R)$, they won't be enough to lift the zero modes $\chi^1_L$ and $\psi^2_R$. 
As we will see below, only a non-zero VEV for the $A_5$-Higgs can rotate the multiplets and lift the two zero modes.
To obtain the matrix of Green's functions we divide the problem into the same domains as in the gauge case, and solve for
$\ola{A}_j,\ola{B}_j,\ora{A}_j,\ora{B}_j,A^\infty_j,B^\infty_j$, where this time $i,j$ are joint indices in $\{1L,2L,1R,2R\}$.

The boundary and jump conditions in this case are:
\begin{itemize}
\item UV BC:
\begin{eqnarray}
G^{\psi}_{L1;j}|_{z=R}~=~G^{\chi}_{L2;j}|_{z=R}~=~G^{\psi}_{R1;j}|_{z=R}~=~G^{\chi}_{R2;j}|_{z=R}~=~0\, .
\end{eqnarray}
\item Jump conditions at $z=z'$:
\begin{eqnarray}
\Delta G^\psi_{ij}|_{z=z'}~=~-\Delta G^\chi_{ij}|_{z=z'}~=~D_{ij}\,a(z')^{-4}\, .
\end{eqnarray}
\item Jump conditions at $z=R'$:
On the IR brane we turn on the masses $M_1 \,\bar{Q}^1_L\, {Q}^1_R$ and $M_2 \,\bar{Q}^2_L\, {Q}^2_R$, which give us the jump condition:
\begin{equation} \label{eq:fIR2} \begin{split}
\Delta G^\chi_{ak;j}|_{z=R'}~&=~-M_k\left<G^\chi_{\bar{a}k;j}\right>|_{z=R'}\, \\
\Delta G^\psi_{ak;j}|_{z=R'}~&=~~~~~M_k\left<G^\psi_{\bar{a}k;j}\right>|_{z=R'} \,,
\end{split} \end{equation}
where $j\in\{1L,2L,1R,2R\}$, $a\in\{L,R\}$, and $k\in\{1,2\}$.
\item Regularity of $G(z,z;p^2)$ at $z\rightarrow \infty$: as before, $A^\infty_{ij}=0$.
\end{itemize}

Note that we have yet to turn on a Higgs VEV mixing the $1$ and $2$ states. 
At this stage there are zero modes in $\chi^1_L$ and in $\psi^2_R$. 
The IR masses $M_{1,2}$ cannot change this fact, we can only lift the two zero modes with a mass that connects them to each other. 
This is impossible as long as the $1$ and $2$ BC are completely decoupled. 
This is not the case when a Higgs VEV is turned on. As usual, this VEV appears as a rotation of the IR jump conditions:
\begin{equation} \begin{split}
\hat{G}_{a1;j}|_{z=R'}~&=~~~~c_h\,G_{a1;j}~+~s_h\,G_{a2;j}\,, \\
\hat{G}_{a2;j}|_{z=R'}~&=~-s_h\,G_{a1;j}~+~c_h\,G_{a2;j} \,.
\end{split} \end{equation}
The rotation is the same for $G^\chi_{ij}$ and $G^\psi_{ij}$. 
We now write the IR BC in Eq.~\ref{eq:fIR2} in terms of the \textit{Higgs-rotated} Green's functions $\hat{G}_{ij}$.

\begin{figure}
\begin{center}
\includegraphics[width=0.6\textwidth]{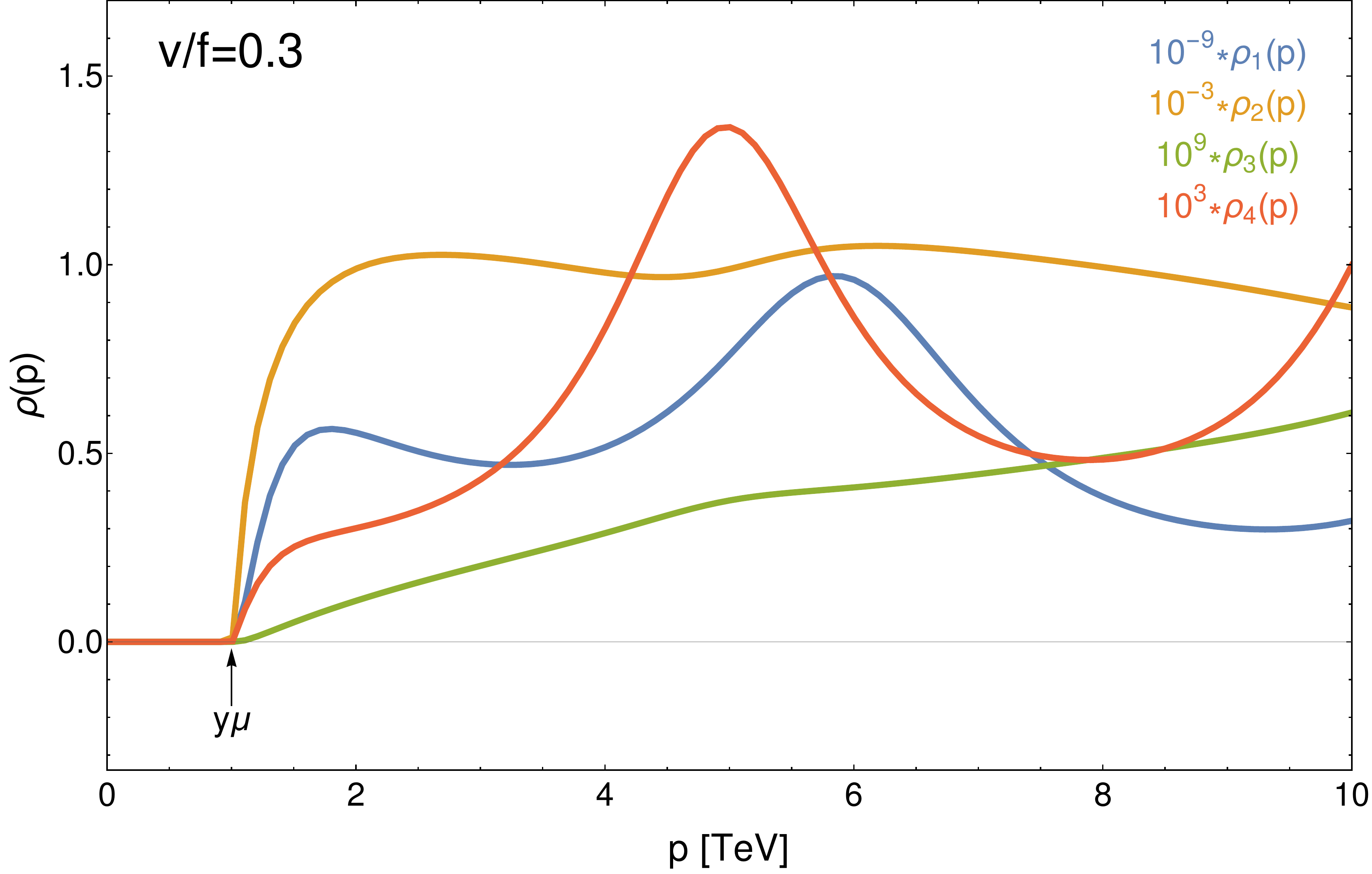}
\end{center}
\caption{Spectral density for $c_L=0.3,\,c_R=-0.1,\,M_1=0.6,\,M_2=0$.}
\label{fig:rhoh}
\end{figure}

\begin{figure}
\begin{center}
\includegraphics[width=0.6\textwidth]{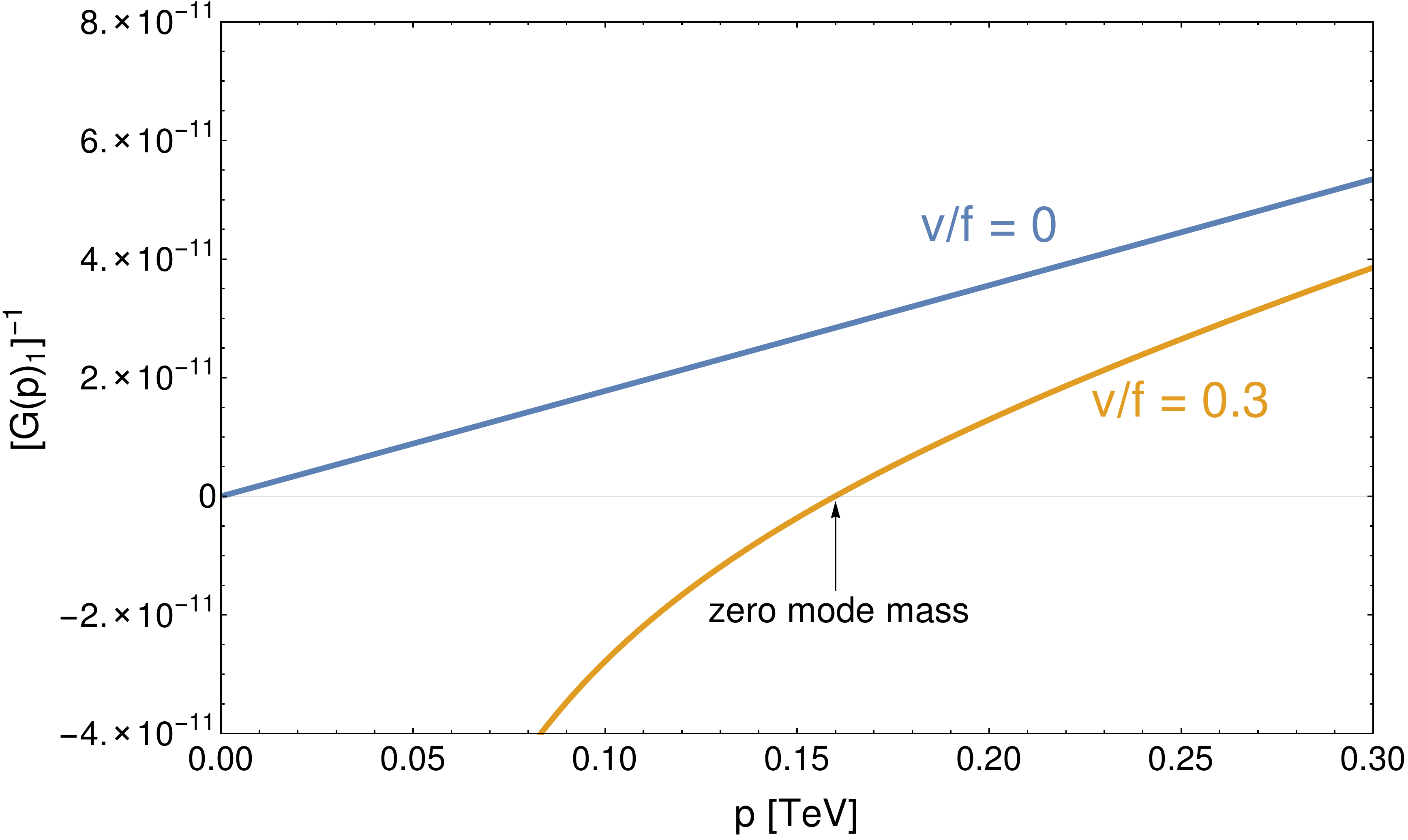}
\end{center}
\caption{Inverse Green's function for $c_L=0.3,\,c_R=-0.1,\,M_1=0.6,\,M_2=0$. The inverse Green's function gets a non-trivial zero for $v/f=0.3$.}
\label{fig:irhoh}
\end{figure}

Solving these conditions, we obtain the Green function $G^h_{ij}$ with $G_{a1;j}=G^\chi_{a1;j}~,~G_{a2;j}=G^\psi_{a2;j}$. 
This matrix now has four eigenvalues $G^h(z,z';p)_{1,2,3,4}$ with appropriate spectral densities $\rho^h(p)_{1,2,3,4}$. 
The label $h$ is an explicit reminder that they depend on the Higgs VEV. 
In Fig.~\ref{fig:rhoh} we plot the spectral densities for $R'={\left(4\,\text{TeV}\right)}^{-1},\,M_1=0.3,\, M_2=0,\,c_L=0.3$ and $c_R=-0.1$, and $v/f=0.3$. 
In Fig.~\ref{fig:irhoh} we plot the inverse Green's functions, and show that there are non-trivial zeros for $v/f=0.3$.

We are now ready to present the spectral densities for the top/bottom sector for the realistic model. 
We will choose the fermions to be embedded in the bulk multiplets $Q_L,\,T_R$ and $B_R$ in the $\mathbf{5}_{\frac{2}{3}},\mathbf{5}_{\frac{2}{3}}$ and $\mathbf{10}_{\frac{2}{3}}$ of $SO(5)\times U(1)_X$. 
The SM gauge group $SU(2)_L$ is a subgroup of $SO(5)$, and the hypercharge is a combination of $U(1)_R\subset SO(5)$ and $U(1)_X$ defined by $Y=T^3_R+X$. 
Under the subgroup $SU(2)_L\times U(1)_Y$, the bulk multiplets decompose as:
\begin{equation} \begin{split}
Q_L(\mathbf{5})_{\frac{2}{3}}~&\rightarrow~q_L(\mathbf{2})_{\frac{1}{6}}~+~\tilde{q}_L(\mathbf{2})_{\frac{7}{6}}~+~y_L(\mathbf{1})_{\frac{2}{3}}\,, \\
T_R(\mathbf{5})_{\frac{2}{3}}~&\rightarrow~q_R(\mathbf{2})_{\frac{1}{6}}~+~\tilde{q}_R(\mathbf{2})_{\frac{7}{6}}~+~t_R(\mathbf{1})_{\frac{2}{3}}\,, \\
B_R(\mathbf{10})_{\frac{2}{3}}~&\rightarrow~q'_R(\mathbf{2})_{\frac{1}{6}}~+~\tilde{q}'_R(\mathbf{2})_{\frac{7}{6}}~+~x_R(\mathbf{3})_{\frac{2}{3}}~+~y_R(\mathbf{1})_{\frac{7}{6}}~+~\tilde{y}_R(\mathbf{1})_{\frac{1}{6}}~+~b_R(\mathbf{1})_{-\frac{1}{3}}\, .
\end{split} \end{equation}
We choose the UV BC such that only $q_L,\,t_R$ and $b_R$ have zero modes by assigning Neumann BC for $G^\chi_{q_L},\,G^\psi_{q_R}$ and $G^\psi_{B_R}$, and Dirichlet BC for all the other $G^\chi_{Q_L},\,G^\psi_{T_R}$ and $G^\psi_{B_R}$.
The $SO(4)\times U(1)_X$ symmetry on the IR brane allows for three mass terms:
\begin{eqnarray}
S_{\text{IR}}~=~\int~dx^4~\sqrt{g_{\text{ind}}}~\left[\,M_1 \, \bar{y}_L t_R~+~M_4 \, \left(\bar{q}_L q_R + \bar{\tilde{q}}_L \tilde{q}_R\right)~+~M_b\,  \left(\bar{q}'_L q_R + \bar{\tilde{q}}'_L \tilde{q}_R\right)\,\right]\, .
\end{eqnarray}
These mass terms lead as usual to IR jump conditions which are for $Q_L$:
\begin{equation} \begin{split} \label{eq:fIRQ}
\Delta G^\chi_{q_L,\tilde{q}_L;j}|_{z=R'}~&=~-M_4\left<G^\chi_{q_R,\tilde{q}_R;j}\right>~-~M_b\left<G^\chi_{q'_R,\tilde{q}'_R;j}\right>|_{z=R'}\,,\\
\Delta G^\psi_{q_L,\tilde{q}_L;j}|_{z=R'}~&=~~~~~M_4\left<G^\psi_{q_R,\tilde{q}_R;j}\right>~-~M_b\left<G^\psi_{q'_R,\tilde{q}'_R;j}\right>|_{z=R'}\,,\\
\Delta G^\chi_{y_L;j}|_{z=R'}~&=~-M_1\left<G^\chi_{t_R;j}\right>|_{z=R'}\,,\\
\Delta G^\psi_{y_L;j}|_{z=R'}~&=~~~~~M_1\left<G^\psi_{t_R;j}\right>|_{z=R'}\, .
\end{split} \end{equation}
For $T_R$ they are:
\begin{equation} \begin{split} \label{eq:fIRT}
\Delta G^\chi_{q_R,\tilde{q}_R;j}|_{z=R'}~&=~-M_4\left<G^\chi_{q_L,\tilde{q}_L;j}\right>|_{z=R'}\,,\\
\Delta G^\psi_{q_R,\tilde{q}_R;j}|_{z=R'}~&=~~~~~M_4\left<G^\psi_{q_L,\tilde{q}_L;j}\right>|_{z=R'}\,,\\
\Delta G^\chi_{t_R;j}|_{z=R'}~&=~-M_1\left<G^\chi_{s_L;j}\right>|_{z=R'}\,,\\
\Delta G^\psi_{t_R;j}|_{z=R'}~&=~~~~~M_1\left<G^\psi_{s_L;j}\right>|_{z=R'}\, , 
\end{split} \end{equation}
and for $B_R$ they are:
\begin{equation} \begin{split} \label{eq:fIRB}
\Delta G^\chi_{q'_R,\tilde{q}'_R;j}|_{z=R'}~&=~-M_b\left<G^\chi_{q_L,\tilde{q}_L;j}\right>|_{z=R'}\,,\\
\Delta G^\psi_{q'_R,\tilde{q}'_R;j}|_{z=R'}~&=~~~~~M_b\left<G^\psi_{q_L,\tilde{q}_L;j}\right>|_{z=R'}\,,\\
\Delta G^\chi_{x_R,y_R,\tilde{y}_R,b_R;j}|_{z=R'}~&=~~~~~~~~~~~~~~~0\,,\\
\Delta G^\psi_{x_R,y_R,\tilde{y}_R,b_R;j}|_{z=R'}~&=~~~~~~~~~~~~~~~0\,.
\end{split} \end{equation}

As usual, in the IR jump conditions, the fields to the left of the IR brane are really the Higgs-rotated fields:
\begin{eqnarray}
\hat{G}_{Q_L}~=~U_h\,G_{Q_L}~,~\hat{G}_{T_R}~=~U_h\,G_{T_R}~,~\hat{G}_{B_R}~=~U_h\,G_{B_R}\,U^{-1}_h\, ,
\end{eqnarray}
decomposed into their different components.
The jump discontinuity due to IR localized mass terms results in quasi-IR brane-localized wave function profiles for the fermionic fields as noted at the end of \secref{fullmod}.

\end{appendix}

\bibliographystyle{JHEP}
\bibliography{continuum}{}

\end{document}